\documentclass[a4paper,11pt,bbold,amssymb,amsmath]{article}
\pdfoutput=1 

\usepackage{jheppub}
\usepackage{slashed}
\usepackage{amsthm}
\usepackage{wrapfig}
\usepackage{color}
\usepackage{bm}

\definecolor{orange}{rgb}{1,0.5,0}
\definecolor{dark-green}{rgb}{0.,0.5,0.0}

\newcommand{\tE}{{\widetilde E}}

\newcommand{\sign}[1]{{\mathrm{sign}}(#1)}
\newcommand{\beqn}{\begin{eqnarray}}
\newcommand{\eeqn}{\end{eqnarray}}
\newcommand{\eq}[1]{(\ref{#1})}
\newcommand{\jj}{{\mathrm{j}}}

\newcommand{\cL}{{\cal L}}
\newcommand{\cZ}{{\cal Z}}

\newcommand{\lab}{{\mathrm {lab}}}

\newcommand{\Z}{{\mathbb Z}}
\newcommand{\R}{{\mathbb R}}

\newcommand{\bs}{\boldsymbol}

\newcommand{\avr}[1]{{\left\langle #1 \right\rangle}}

\title{
Interacting fermions in rotation: chiral symmetry restoration, moment of inertia and thermodynamics
}

\author[a,b]{M. N. Chernodub}
\author[a,c]{and Shinya Gongyo}
\affiliation[a]{CNRS, Laboratoire de Math\'ematiques et Physique Th\'eorique, Universit\'e de Tours, France}
\affiliation[b]{Laboratory of Physics of Living Matter, Far Eastern Federal University, Vladivostok, Russia}
\affiliation[c]{Theoretical Research Division, Nishina Center, RIKEN, Saitama, Japan}

\abstract{We study rotating fermionic matter at finite temperature in the framework of the Nambu--Jona-Lasinio model. In order to respect causality the rigidly rotating system must be bound by a cylindrical boundary with appropriate boundary conditions that confine the fermions inside the cylinder. We show the finite geometry with the MIT boundary conditions affects strongly the phase structure of the model leading to three distinct regions characterized by explicitly broken (gapped), partially restored (nearly gapless) and spontaneously broken (gapped) phases at, respectively, small, moderate and large radius of the cylinder. The presence of the boundary leads to specific steplike irregularities of the chiral condensate as functions of coupling constant, temperature and angular frequency. These steplike features have the same nature as the Shubnikov--de Haas oscillations with the crucial difference that they occur in the absence of both external magnetic field and Fermi surface. At finite temperature the rotation leads to restoration of spontaneously broken chiral symmetry while the vacuum at zero temperature is insensitive to rotation (``cold vacuum cannot rotate''). As the temperature increases the critical angular frequency decreases and the transition becomes softer.  A phase diagram in angular frequency-temperature plane is presented. We also show that at fixed temperature the fermion matter in the chirally restored (gapless) phase has a higher moment of inertia compared to the one in the chirally broken (gapped) phase. }

\begin{document} 

\maketitle
\flushbottom

\section{Introduction}

Recently much interest has been attracted to rigidly rotating systems of interacting relativistic fermions. At the particle physics side the interest is heated by the fact that the noncentral heavy-ion collisions produce rapidly rotating quark-gluon plasma which should carry large global angular momentum~\cite{ref:HIC:1,ref:HIC:2,ref:HIC:3,ref:HIC:4}. The rotating relativistic systems experience anomalous transport phenomena such as, for example, the chiral vortical effect~\cite{ref:CVE:2} which also has been discussed in an astrophysical context earlier~\cite{ref:Vilenkin}. In the condensed matter the relativistic fermions appear in Weyl/Dirac semimetals that may also be sensitive to rotation due to the anomalous transport~\cite{ref:Weyl:1,ref:Weyl:2,ref:Weyl:3}. 

Theoretically, the problem of rotating free fermion states has recently been addressed in the Refs.~\cite{Ambrus:2014uqa,Ambrus:2015lfr} and the system of interacting fermions has been studied both in unbounded~\cite{Chen:2015hfc,Jiang:2016wvv} and bounded~\cite{Ebihara:2016fwa} geometries in effective field-theoretical models, as well as in the holographic approaches~\cite{McInnes:2014haa,McInnes:2015kec,McInnes:2016dwk}. Properties of rotating strongly interacting matter were also probed in lattice simulations of euclidean QCD~\cite{Yamamoto:2013zwa}. 

In our paper we study how rotation affects phase structure and thermodynamics of a system of interacting fermions described by the Nambu--Jona-Lasinio model~\cite{ref:NJL}. We point out that a rigidly rotating system must be bounded in directions transverse to the axis to rotation in order to avoid the causality violation. The unbounded rotating systems may have several pathologies related to instabilities and excitations by rotation when a region of the rotating space exceeds the speed of light~\cite{Ambrus:2014uqa,ref:Levin,Davies:1996ks}. Thus the rotating space should be bounded which immediately implies that the physics of the system must be dependent on the type of the boundary conditions that are imposed in the finite space directions. The importance of the boundary effects has also been recently noted in a different approach in Ref.~\cite{Ebihara:2016fwa}. In our study we consider a system of rigidly rotating fermions in a region bounded by a cylindrical shell at which the fermions are subjected to the MIT boundary conditions. 

The structure of this paper is as follows. First, in Sect.~\ref{sec:free:fermions} we discuss free massive rigidly rotating fermions in a cylinder of a finite radius: in Sect.~\ref{sec:spectrum} we review the details of the spectrum obtained in Ref.~\cite{Ambrus:2015lfr}, then in Sect.~\ref{sec:structure} we describe specific features of the structure of the energy spectrum and, based on these results, we discuss similarities and differences between global rotation and external magnetic field in Sect.~\ref{sec:rotation:vs:magnetic}. Our study supports the idea that the properties of the energy spectrum of the relativistic rotating fermions imply that the rotation in a relativistic system cannot be associated with the presence of an effective magnetic field background. 

\begin{wrapfigure}{r}{0.45\textwidth}
\vskip -3mm
\centerline{\includegraphics[scale=0.6,clip=true]{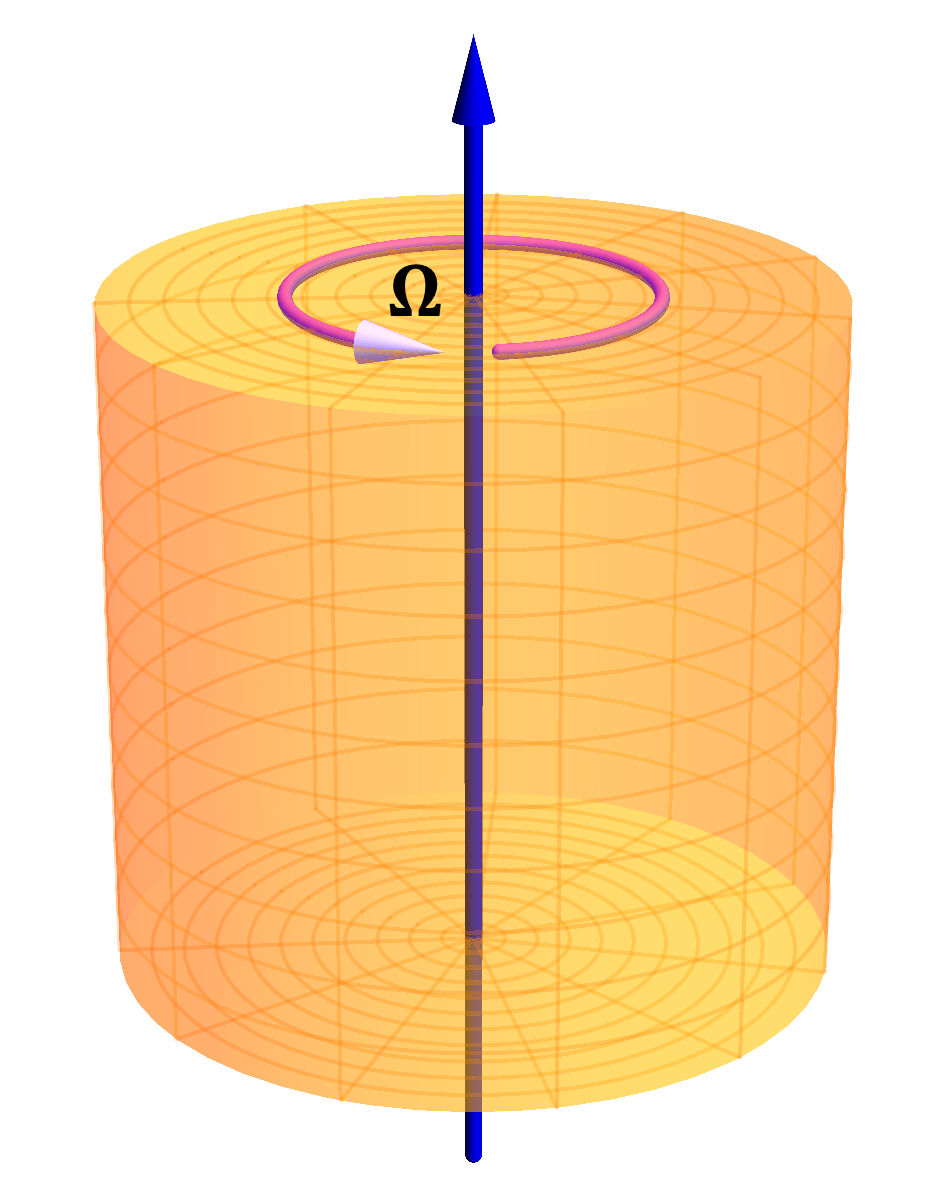}}
\caption{The fermionic medium is uniformly rotating with constant angular velocity $\Omega$ inside the cylinder of fixed radius $R$.}
\label{fig:cylinder}
\end{wrapfigure}

In Section~\ref{sec:interacting} we describe rotating interacting fermions in the framework of the Nambu--Jona-Lasinio model. We stress that a rigid rotation of a relativistic system in thermal equilibrium is necessarily a finite-geometry problem. The phase diagram of the system at zero temperature is calculated in Sect.~\ref{sec:T:0:phase}. In Section~\ref{sec:cold:vacuum} we demonstrate that the cold vacuum cannot rotate. The properties of the rotating fermionic matter at finite temperature are discussed in details in Sect.~\ref{sec:rotating:finite:T}. We find the finite-temperature phase diagram in Sect.~\ref{sec:phase}, discuss the angular momentum and moment of inertia in different phases (Sect.~\ref{sec:angular}), and calculate energy and entropy densities of rotating fermions (Sect.~\ref{sec:energy}). The last Section is devoted to discussions and conclusions. 

\section{Free massive fermions in rotation}
\label{sec:free:fermions}

\subsection{Spectrum of rigidly rotating fermions inside a cylinder}
\label{sec:spectrum}

In this Section we describe certain properties of solutions to the Dirac equation for uniformly rotating massive fermions inside a cylindrical cavity, Fig.~\ref{fig:cylinder}, following closely Ref.~\cite{Ambrus:2015lfr}. 

We consider a fermionic system which uniformly rotates with the constant angular velocity $\Omega$ about the fixed $z \equiv x_3$ axis. We assume that all spatial regions of the system have the same angular velocity so that the rotation is rigid. The rigid nature of rotation immediately implies that the system must have a finite size in the plane perpendicular to the axis of rotation. Indeed, the absolute value of velocity of a point located at the distance $\rho$ from the axis of rotation, $v = \Omega \rho$ should not exceed the speed of light to preserve the causality $\rho \Omega \leqslant 1$. Without loss of generality we assume that the rotation is always going into the counterclockwise direction so that $\Omega \geqslant 0$ throughout all the paper.

Due to the symmetry of the problem it is convenient to consider rotating volumes with cylindrical geometries, Fig.~\ref{fig:cylinder}. It is natural to choose the cylindrical coordinates, $x \equiv (x_0, x_1,x_2,x_3) = (t,\rho\sin\varphi,\rho\cos\varphi,z)$. The coordinates $t$, $\rho$ and $z$ in the corotating reference frame (which rotates together with the system) coincide with the corresponding coordinates of the laboratory frame: $t = t_{\lab}$, $\rho = \rho_\lab$ and $z = z_\lab$. The angular variables in these frames are related as follows: 
\beqn
\varphi = [\varphi_{\mathrm{lab}} - \Omega t]_{2\pi}\,,
\label{eq:varphi}
\eeqn
where $[\dots]_{2\pi}$ means ``modulo $2\pi$''. In both frames the boundary of the cylindric volume is given by $\rho = R$. The requirement of causality implies that product of the angular velocity and the radius of rigidly rotating cylinder is bound:
\beqn
\Omega R \leqslant 1\,.
\label{ref:bound}
\eeqn

In order to preserve the number of fermions inside the cylindrical cavity it is natural to impose on the fermion wavefunctions the MIT conditions at the boundary of the cylinder~$\rho = R$,
\beqn
\bigl[i \gamma^\mu n_\mu(\varphi) - 1 \bigr] \psi(t,z,\rho,\varphi) {\biggl |}_{\rho = R} = 0\,, \qquad \qquad \mbox{[MIT b.c.]}
\label{eq:MIT:boundary}
\eeqn
where $n_\mu(\varphi) = (0, R \cos\varphi, - R \sin\varphi, 0)$ is a vector normal to the cylinder surface and $\gamma^\mu$ are the Dirac matrices. The boundary condition~\eq{eq:MIT:boundary} confines the fermions inside the cavity because Eq.~\eq{eq:MIT:boundary} forces the normal component of the fermionic current,
\beqn
j^\mu = {\bar \psi} \gamma^\mu \psi\,,
\label{eq:j:mu:basic}
\eeqn
to vanish at the surface of the cylinder ($\rho = R$):
\beqn
j_{\bs n} \equiv  {\bs j} {\bs n} \equiv - j^\mu n_\mu 
= 0 \qquad \mbox{at} \quad \rho = R \,.
\label{eq:j:n}
\eeqn

The rotating frame~\eq{eq:varphi} is a curvilinear reference frame with the metric
\begin{align}
g_{\mu \nu} =
\begin{pmatrix}
1-(x^2+y^2)\Omega ^2 & y\Omega & -x\Omega & 0 \\
y\Omega & -1 & 0 & 0 \\
-x\Omega & 0 & -1 & 0 \\
0 & 0 & 0 &-1
\end{pmatrix}
,
\end{align}
which corresponds to the line element
\beqn
ds^2 \equiv g_{\mu\nu} dx^\mu dx^\nu = \left(1-\rho ^2 \Omega ^2 \right)dt^2- 2\rho^2\Omega dt d\varphi - d\rho ^2- \rho^2 d\varphi^2 - d z^2\,.
\label{eq:metric}
\eeqn
Here we adopt the convention that $\hat{i},\hat{j} \dots = \hat{t},\hat{x},\hat{y},\hat{z}$ and $\mu,\nu \dots = t, x, y, z$ refer to the Cartesian coordinate in the local rest frame, corresponding to the laboratory frame, and the general coordinate in the rotating frame, respectively.

Consequently, a fermion with the mass $M$ is described by the Dirac equation in the curved spacetime
\beqn
\left[ i \gamma^\mu \left(\partial_\mu + \Gamma^\mu \right)- M \right] \psi = 0\,.
\label{eq:Dirac}
\eeqn
where the affine connection $\Gamma _\mu$ is defined by
\begin{align}
\Gamma_\mu = -\frac{i}{4}\omega_{\mu i j}\sigma^{ij}, 
\qquad
\omega_{\mu \hat{i}\hat{j}} = g_{\alpha \beta} e^{\alpha}_{\hat{i}}\left(\partial_{\mu}e^{\beta}_{\hat{j}}+ \Gamma^{\beta}_{\nu \mu}e^{\mu}_{\hat{j}}\right), 
\qquad
\sigma^{\hat{i}\hat{j}}  = \frac{i}{2}\left[\gamma^{\hat{i}}, \gamma^{\hat{j}}\right],
\end{align}
with the Christoffel connection, $\Gamma _{\mu \nu} ^{\lambda} = \frac{1}{2}g^{\lambda\sigma}\left(g_{\sigma \nu,\mu}+g_{\mu\sigma,\nu}-g_{\mu\nu,\sigma}\right)$, and the gamma matrix in curved space-time, $\gamma ^\mu = e^{\mu} _{\hat{i}} \gamma ^{\hat{i}}$, fulfills the anti-commutation relation,
\begin{align}
\left\{\gamma^\mu, \gamma^\nu\right\} = 2 g^{\mu \nu}.
\end{align}

The vierbein $e^{\mu}_{\hat{i}}$ is written in the Cartesian gauge \cite{Ambrus:2014uqa}, which connects the general coordinate with the Cartesian coordinate in the local rest frame, $x^\mu = e^{\mu}_{\hat{i}}x^{\hat{i}}$, is given by
\begin{align}
e^t_{\hat{t}}=e^x_{\hat{x}}=e^y_{\hat{y}}=e^y_{\hat{y}}=1,
\qquad 
e^x_{\hat{t}}= y\Omega, 
\qquad
e^y_{\hat{t}}= -x\Omega,
\end{align}
and the other components being zero. This leads to the metric $\eta_{\hat{i} \hat{j}} = g_{\mu \nu}e^{\mu}_{\hat{i}}e^{\nu}_{\hat{j}}$.

In the rotating case, the nonzero components of the Christoffel connection are 
\begin{align}
\Gamma_{tx}^{y}=\Gamma_{xt}^{y}= \Omega,~~\Gamma_{ty}^{x}=\Gamma_{yt}^{x}= -\Omega,~~
\Gamma_{tt}^{x}= -x \Omega ^2,~~\Gamma_{tt}^{y}= -y \Omega ^2,
\end{align}
and the other components are zero. Thus, the only nonzero component of $\Gamma _\mu $ is given by
\begin{align}
\Gamma _{t}= -\frac{i}{2} \Omega \, \sigma^{\hat{x}\hat{y}}.
\label{eq:Gamma}
\end{align}
The gamma matrix in the rotating case is given by
\begin{align}
\gamma ^t = \gamma ^{\hat{t}},
\qquad
\gamma^x = y\Omega \gamma^{\hat{t}}+\gamma^{\hat{x}},
\qquad
\gamma^y = -x\Omega \gamma^{\hat{t}}+\gamma^{\hat{y}},
\qquad 
\gamma^z = \gamma^{\hat{z}}.
\end{align}
The Dirac equation is rewritten as
\begin{align}
\left[i\gamma ^{\hat{t}}\left(\partial_t+y\Omega\partial_x -x \Omega \partial_y - \frac{i}{2}\Omega \sigma^{\hat{x}\hat{y}}\right)+ i\gamma ^{\hat{x}}\partial_x+i\gamma^{\hat{y}}\partial_y+ i\gamma^{\hat{z}}\partial_z - M\right]\psi = 0. \label{eq:rotDirac}
\end{align}
In the Dirac representation, 
\begin{align}
\sigma^{\hat{x}\hat{y}}=
\begin{pmatrix}
\sigma^3 & 0 \\
0 & \sigma ^3 
\end{pmatrix}
,
\end{align}
and the Dirac equation is reduced to
\begin{align}
\left[\gamma ^{\hat{t}}\left(i\partial_t+\Omega J_z\right)+ i\gamma ^{\hat{x}}\partial_x+i\gamma^{\hat{y}}\partial_y+ i\gamma^{\hat{z}}\partial_z - M\right]\psi = 0.
\end{align}
with $\hat{J}_z$ being the z-component angular momentum,
\begin{align}
\hat{J}_z  = -i\left(-y\partial_x + x\partial_y \right) + \frac{1}{2}
\begin{pmatrix}
\sigma_3 & 0 \\
0 & \sigma_3 
\end{pmatrix} 
\equiv -i\partial_\varphi + \frac{1}{2}
    \begin{pmatrix}
\sigma_3 & 0 \\
0 & \sigma_3 
\end{pmatrix}
.
\label{eq:hat:J}
\end{align}

Alternatively, the Dirac equation in the rotating reference frame may also be obtained by shifting the free Hamiltonian in the laboratory frame as follows
\begin{align}
H_{\mathrm{lab}}= \int d^3x_{\mathrm{lab}} \mathcal{H}_{\mathrm{lab}}\left(\psi(x_{\mathrm{lab}}) ,\psi^\dagger (x_{\mathrm{lab}}) \right) \rightarrow H_{\mathrm{rot}} \equiv \int d^3x \mathcal{H}_{\mathrm{lab}}\left(\psi(x) ,\psi^\dagger (x) \right)  - \Omega \mathcal{J}_3,
\label{eq:H:rot:shift}
\end{align}
where $\mathcal{J}_3$ is the $z$--component of the fermionic angular momentum operator: 
\begin{align}
\mathcal{J}_3 = \mathcal{M}_{12} \equiv  \int d^3x \psi^ \dagger (x)\left[i\left(-x\partial_y+y\partial_x\right)+\frac{1}{2}\sigma^{\hat{x}\hat{y}}\right]\psi (x)\,.
\end{align}

The Hamiltonian in the rotating reference frame~\eq{eq:H:rot:shift} is thus given by
\begin{align}
H_{\mathrm{rot}}
&=\int d^3x \psi ^\dagger\left[ \left(-i \alpha^{\hat{i}}\partial_i + M\beta \right) -\Omega\left(i\left(-x\partial_y+y\partial_x\right)+\frac{1}{2}\sigma^{\hat{x}\hat{y}}\right)\right] \psi 
\end{align}
with $\alpha^{\hat{i}} = \gamma^0 \gamma ^{\hat{i}}$, and $\beta = \gamma ^0,$ and $\alpha^{\hat{i}}\partial_i=\alpha^{\hat{x}}\partial_x+\alpha^{\hat{y}}\partial_y+\alpha^{\hat{z}}\partial_z$. The Dirac equation~\eq{eq:rotDirac} is obtained as a Heisenberg equation, using the equal-time anticommutation relation obtained from the canonical quantization.

A general solution of the Dirac equation~\eq{eq:rotDirac} with the boundary conditions~\eq{eq:MIT:boundary} in the rotating reference frame~\eq{eq:varphi} has the following form:
\beqn
U_j = \frac{1}{2\pi} e^{- i \tE t + i k_z z} u_j(\rho,\varphi)\,,
\eeqn
where $u_j$ is an eigenspinor characterized by the set of quantum numbers,
\beqn
j = (k_z, m, l, \sign{E})\,, \qquad m \in \Z\,, \qquad l = 1,2,\dots\,, \qquad k_z \in \R\,,
\label{eq:j}
\eeqn
$k_z$ is the momentum of the fermion along the $z$ axis, $m$ is the quantized angular momentum with respect to the $z$ axis, and $l$ is the radial quantum number which describes the behavior of the solution in terms of the radial $\rho$ coordinate.

The energy $\tE_j$ in the corotating frame,
\beqn
\tE_j = E_j - \Omega \Bigl(m + \frac{1}{2}\Bigr) \equiv E_j - \Omega \mu_m\,,
\label{eq:Energy}
\eeqn
is related to the energy $E_j$ in the laboratory frame:
\beqn
E_j \equiv E_{ml}(k_z,M) = \pm \sqrt{k_z^2 + \frac{q_{ml}^2}{R^2} + M^2}\,,
\label{eq:E:j}
\eeqn
where the dimensionless quantity $q_{ml}$ is the $l^{\mathrm{th}}$ positive root $(l=1,2, \dots )$ of the following equation:
\beqn
\jj_{m}^2(q) + \frac{2 M R}{q} \,\jj_m(q) -1  = 0\,,
\label{eq:jj}
\label{eq:J}
\eeqn
with 
\beqn
\jj_m(x) = \frac{J_m(x)}{J_{m+1}(x)}\,,
\label{eq:jjm}
\eeqn
and $J_m(x)$ is the Bessel function.

The quantity $\mu_m$ in Eq.~\eq{eq:Energy} is the eigenvalue 
\beqn
{\hat J}_z \psi = \mu_m \psi\,, \qquad \mu_m =  m + \frac{1}{2}\,,
\label{eq:mu:m}
\eeqn
of the $z$-component of the total angular momentum operator which comprises the orbital and spin parts~\eq{eq:hat:J}. Thus the quantiry $\mu_m$ can be identified with the quantized value of the total angular momentum.

In the cylindric volume the integration measure over the momentum is modified with respect to the measure in an unbounded space:
\beqn
\int \frac{d^3 k}{(2 \pi)^3} \to \sum_j \equiv \frac{1}{\pi R^2}\sum_{l=1}^\infty \sum_{m= -\infty}^\infty \int \frac{d k_z}{2 \pi} \,.
\label{eq:phase:space:k}
\eeqn
The integration over continuous 3-momentum is replaced by the integration over the momentum along the axis of rotation $k_z$ and sums over the projection of the angular momentum on the $z$ axis and over the radial excitation number $l$. For a detailed derivation of the energy spectrum we refer the reader to Ref.~\cite{Ambrus:2015lfr}.

\subsection{Structure of the energy spectrum of free rotating fermions}
\label{sec:structure}

The spectrum of rotating fermions contains certain interesting and, sometimes, quite unexpected features.

In the laboratory frame the fermionic eigenenergy~\eq{eq:E:j} does not depend on the angular frequency~$\Omega$. However, in the corotating frame the eigenenergy is a linear function of the angular frequency~$\Omega$. The frequency~$\Omega$ plays the role of a chemical potential associated with the total angular momentum $\mu_m$. 

The change of the orbital number $m \to -m - 1$  does not affect the solutions $q_{m,l}$ of Eq.~\eq{eq:jj} while the total angular momentum $\mu_m$ flips its sign:
\beqn
q_{m,l} = q_{-m-1,l}\,, 
\qquad 
\mu_{m} = - \mu_{-m-1}\,, 
\label{eq:qml:eq}
\eeqn
Therefore the densities of states with positive and negative total angular momenta $\mu_{m}$ are equal to each other. Equation~\eq{eq:qml:eq} also implies that the energy spectrum~\eq{eq:Energy} is double degenerate at $\Omega = 0$.

\begin{figure}[!thb]
\begin{center}
\vskip 5mm
\begin{tabular}{cc}
\includegraphics[scale=0.5,clip=true]{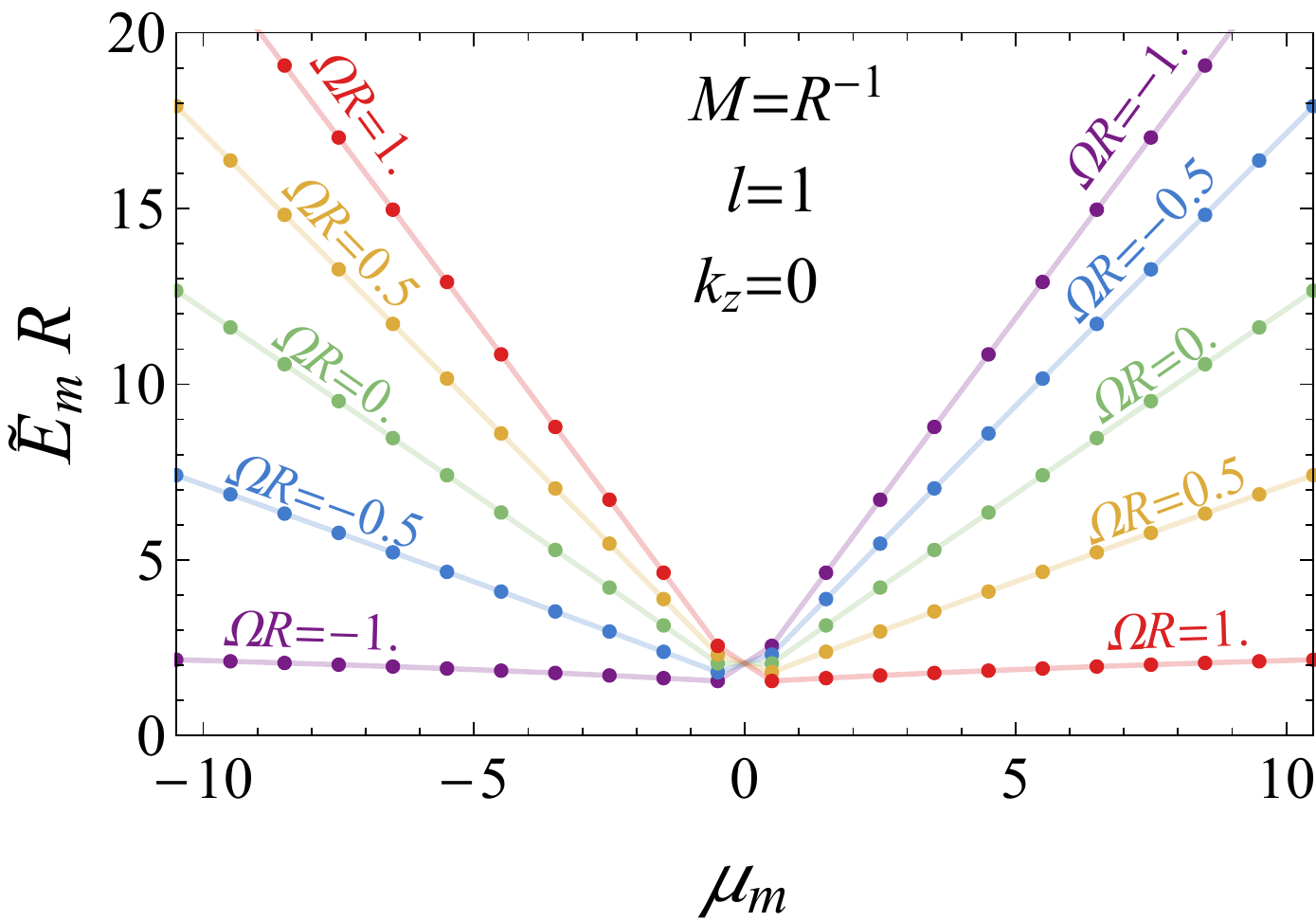} & 
\includegraphics[scale=0.5,clip=true]{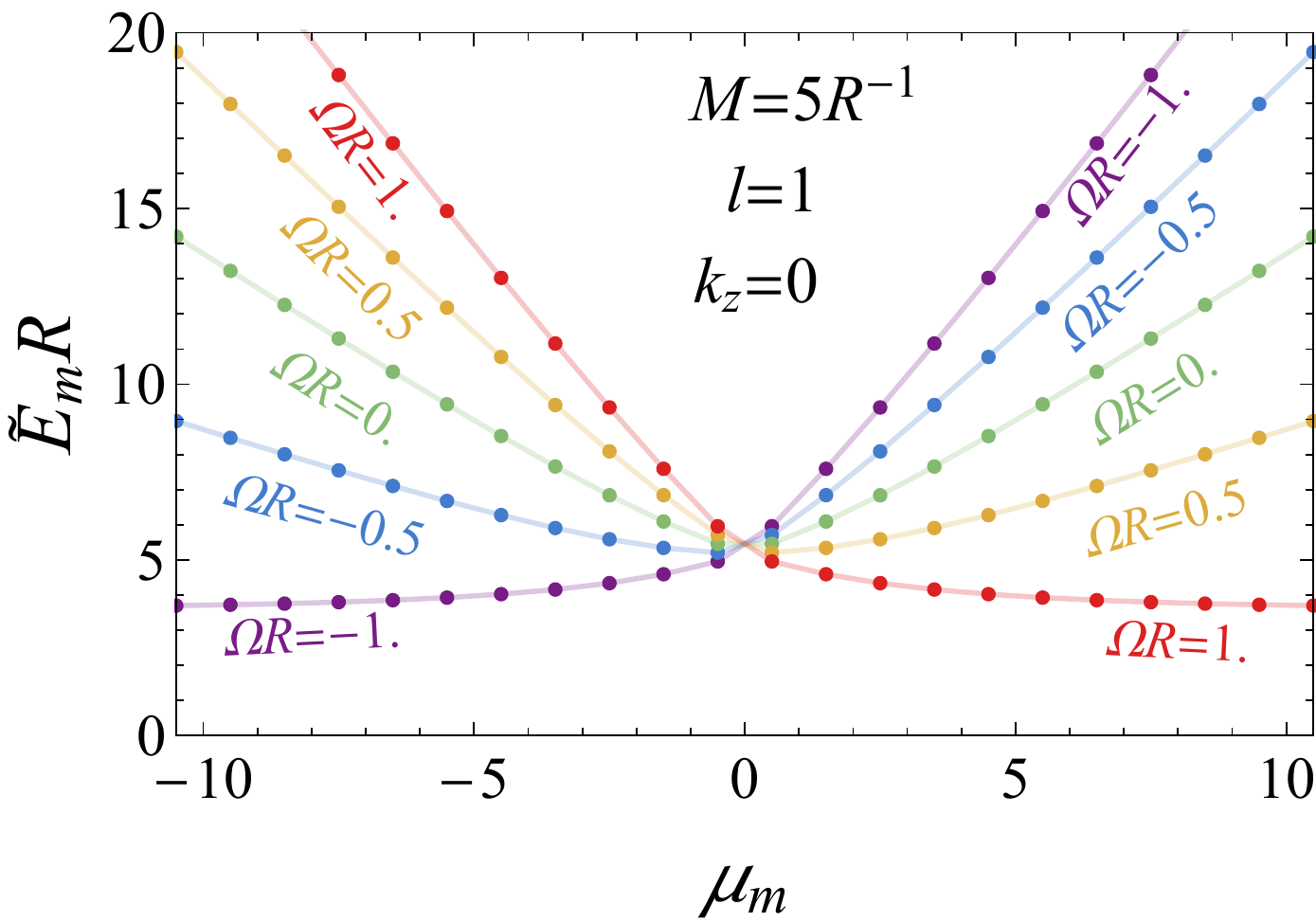}\\
\hskip 8mm (a) & \hskip 8mm (b) \\[3mm]
\includegraphics[scale=0.5,clip=true]{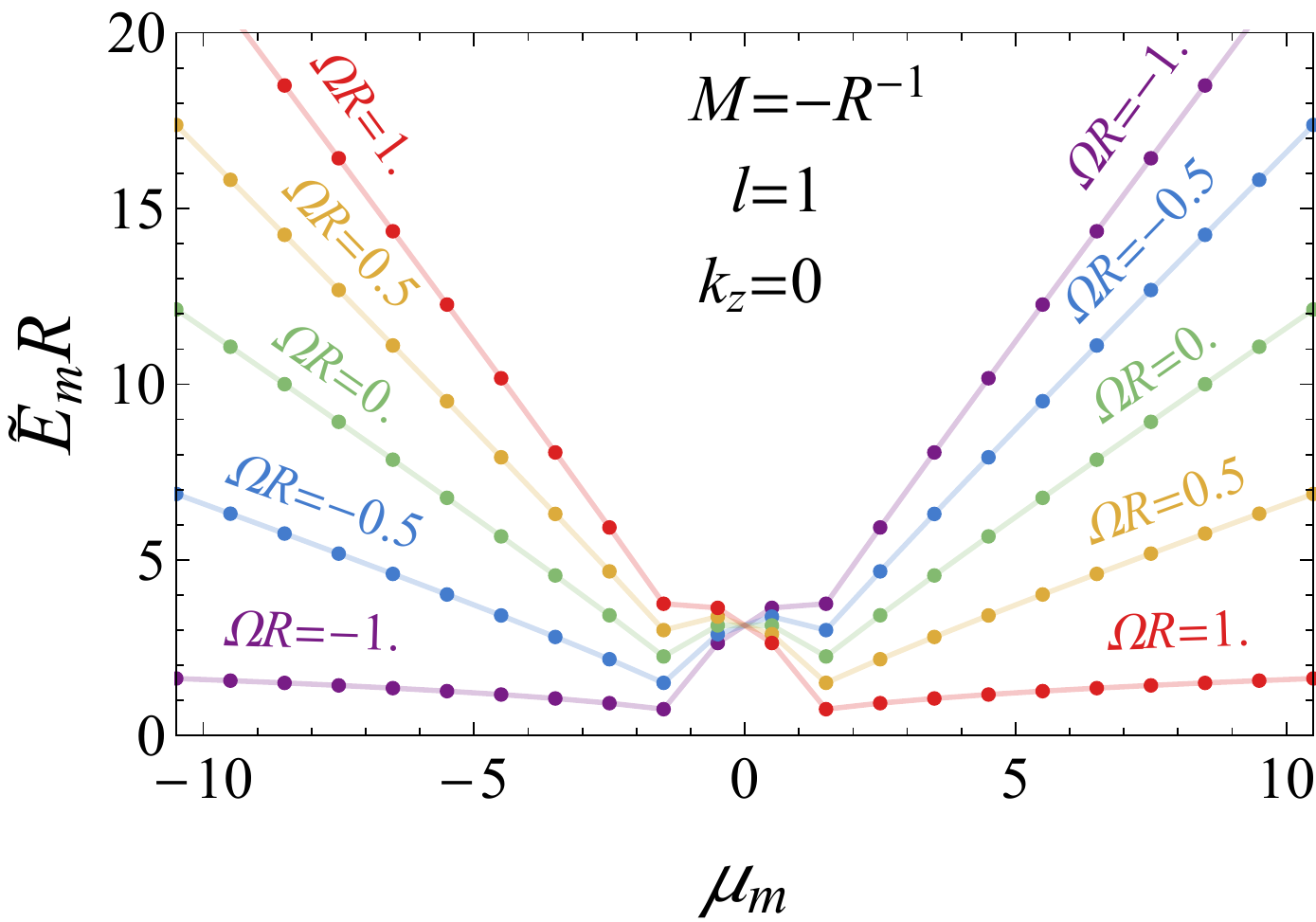} &
\includegraphics[scale=0.5,clip=true]{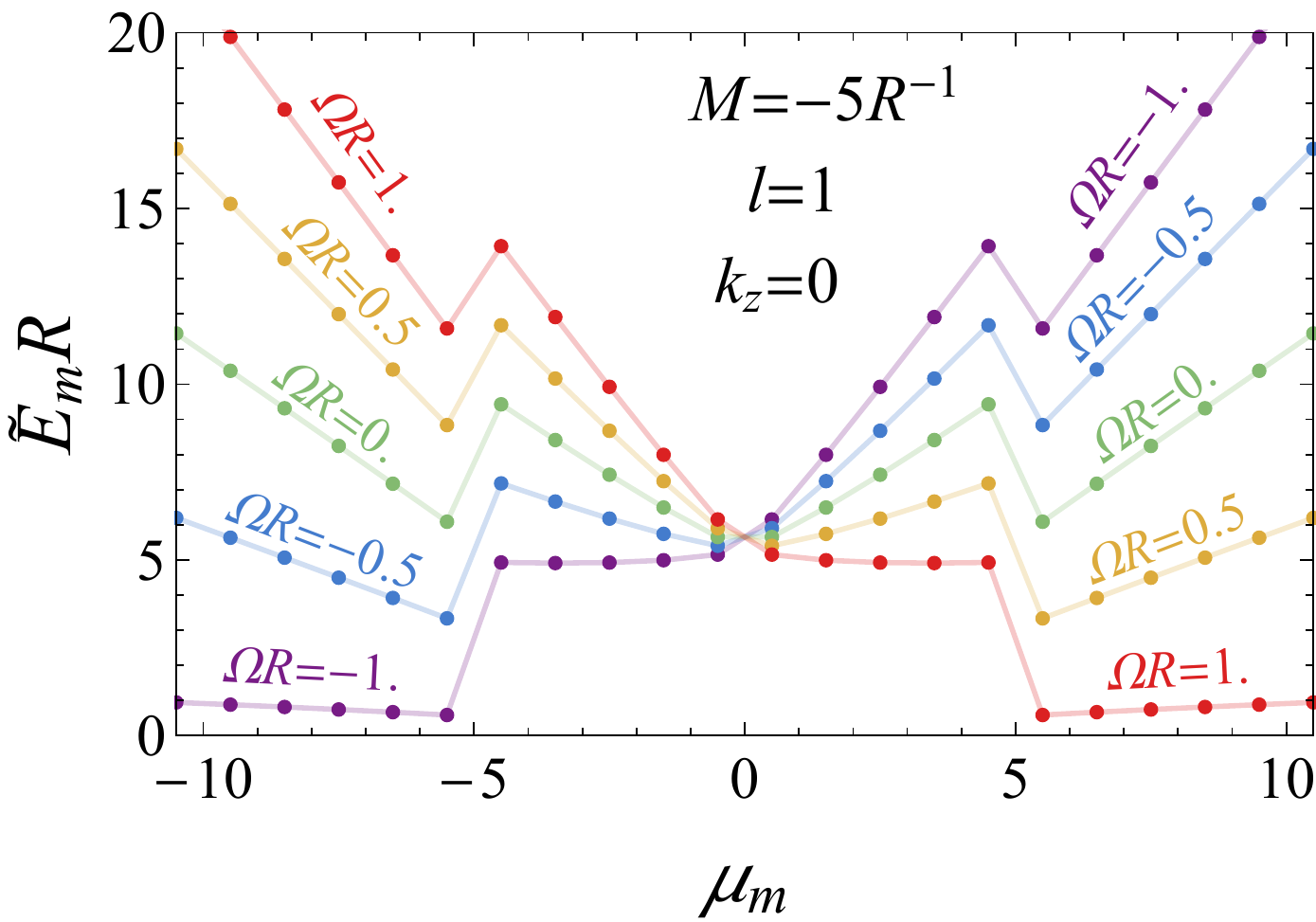}\\
\hskip 8mm (c) & \hskip 8mm (d)
\end{tabular}
\end{center}
\vskip -5mm
\caption{Lowest energy eigenmodes (with $l = 1$ and $k_z = 0$) in the corotating frame~\eq{eq:Energy} vs. the total angular momentum $\mu_m$, Eq.~\eq{eq:mu:m}, for fixed positive (the upper panel) and negative (the lower panel) masses~$M$ for various values of the rotation frequency $\Omega$.}
\label{fig:emum}
\end{figure}

In Fig.~\ref{fig:emum} we show the energy spectrum in the corotating frame as the function of the total angular momentum $\mu_m$ for positive and negative values of masses $M$ and for various values of the angular frequency $\Omega$. We plot only the lowest mode with zero momentum along the rotation axis $k_z = 0$ and with the lowest radial excitation number $l=1$. The modes with $k_z \neq 0$ and higher values of $l \geqslant 2$ lie higher than the corresponding lowest ($k_z = 0$ and $l = 1$) eigenmode and they qualitatively resemble the structure of the lowest eigenmode. We study both positive and negative values of the mass $M$ since in the NJL approach the mass gap is the dynamical variable which may take any real value.

The energy spectra in the corotating frame, for all values of the fermion masses $M$ and for all nonzero angular momenta $\Omega \neq 0$ are asymmetric functions with respect to inversion of the total angular momentum $\mu_m \to -\mu_m$. The asymmetry appears due to the presence of the linear term in the energy in the corotating frame~\eq{eq:Energy}. The presence of the asymmetry indicates that the clockwise and counterclockwise rotations are {\emph {not}} equivalent for each particular mode. However, the spectrum is invariant under the simultaneous flips  $\mu_m \to -\mu_m$ and $\Omega \to -\Omega$. Since the partition function includes the sum over all values of the angular momentum, Eqs.~\eq{eq:Z:NJL:1} and \eq{eq:tr:ln}, then for the whole system the clockwise and counterclockwise rotation are equivalent. For $\Omega = 0$ the spectrum is obviously symmetric under the inversion of the total angular momentum $\mu_m \to -\mu_m$.

At large angular momenta $\mu_m$, the corotating energy spectrum is approaching a linear function. The slopes of this function differ for positive and negative frequencies $\Omega \neq 0$. For positive angular frequencies $\Omega > 0$ the slope of the energy at the positive (negative) angular momenta is milder (steeper) while for negative angular frequencies $\Omega < 0$ the situation is vice versa. One can check numerically that in the limit of large total angular momentum $\mu_m$ the slope of the energy in the rotating frame has the following universal limit ({\emph {cf.}} Fig.~\ref{fig:emum}):
\beqn
\frac{\partial}{\partial \mu_m} {{\tE}_{m,l,k_z}}(\Omega) {\biggl |}_{\mu_m \to \pm \infty} = 1 \mp \Omega R\,.
\label{eq:E:slope}
\eeqn

As one can see from Fig.~\ref{fig:emum}, certain features of the energy spectrum at positive and negative masses $M$ are quite different from each other. 

At positive values of the fermion mass $M>0$ the energy in the corotating frame is always a convex function of the total angular momentum $\mu_m$ (or, equivalently, of the angular momentum number $m$). A change in the value of the mass  does not affect the spectrum qualitatively. Indeed, at low mass the energy spectrum bends sharply at $\mu_m = 0$, Fig.~\ref{fig:emum}(a). As the mass $M$ increases, the behavior of the energy as the function of the total angular momentum $\mu_m$ smoothens at $\mu_m = 0$, Fig.~\ref{fig:emum}(b), while all other features stay the same.  

At negative values of mass, $M < 0$, the energy spectrum becomes more involved. The dependence of energy on the total angular momentum $\mu_m$ becomes generally non-convex. At low negative mass the particularities of the spectrum are observed at $\mu_m \sim 0$, Fig.~\ref{fig:emum}(c), while at large negative mass the spectrum experiences a sharp discontinuity-like feature at the angular momentum $\mu_m \simeq \pm MR$, Fig.~\ref{fig:emum}(d). As we will see below, these features of the energy spectrum will have an interesting effect on the ground state of the system.

\begin{wrapfigure}{r}{0.55\textwidth}
\includegraphics[scale=0.5,clip=true]{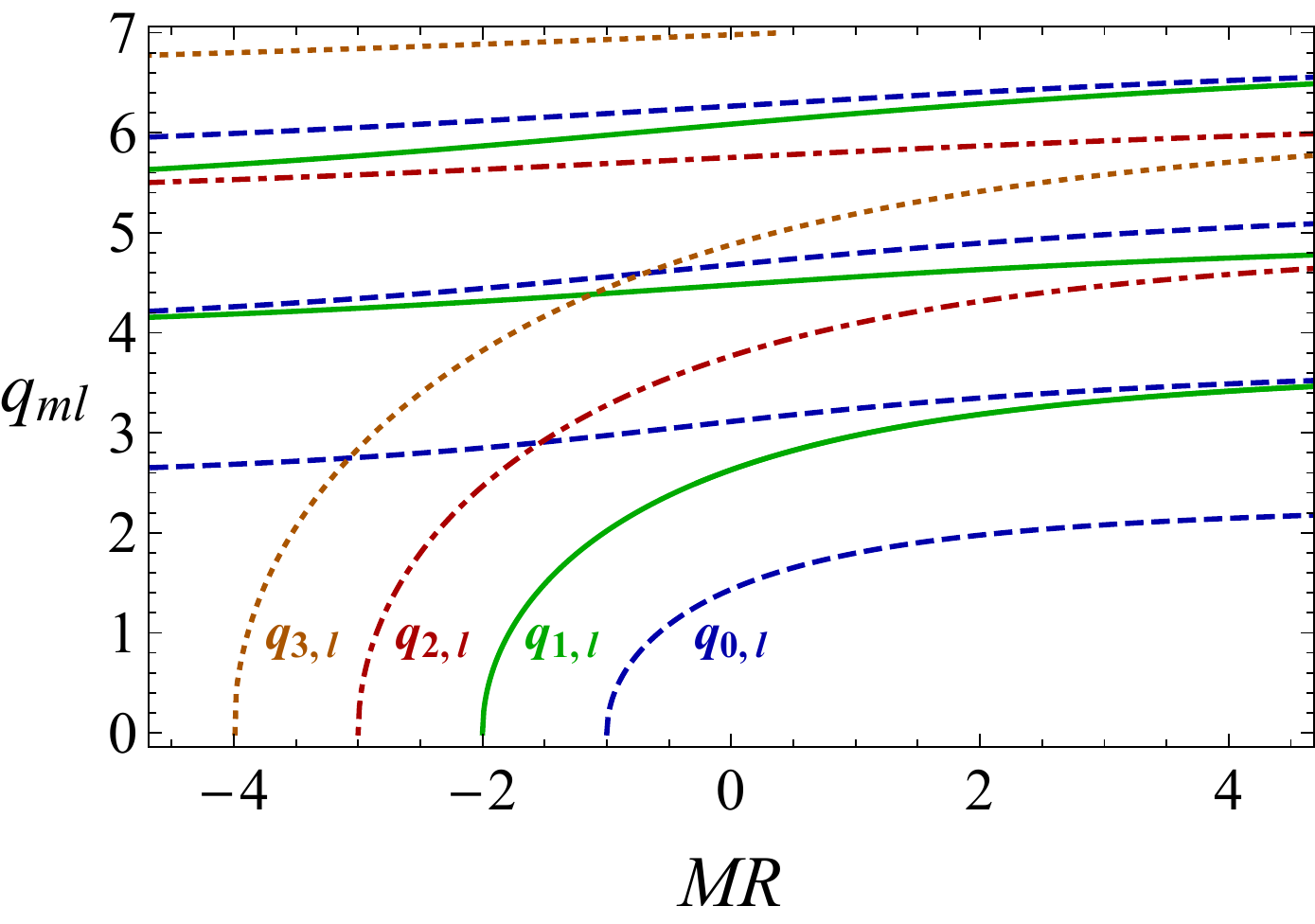}
\vskip -3mm
\caption{Lowest $l = 1,2,\dots$ solutions $q_{ml}$ of Eq.~\eq{eq:J} for the angular quantum numbers $m= 0$ (the blue dashed lines), $m=1$ (the green solid lines), $m=2$ (the red dot-dashed lines) and $m=3$ (the orange dotted lines) vs the normalized mass~$M R$. 
}
\label{fig:q:mls}
\end{wrapfigure}

In order to understand the origin of the discontinuities in the energy spectrum at negative mass $M$ we notice that the solution $q_{ml}$ of Eq.~\eq{eq:J} enters the energy eigenvalue~\eq{eq:E:j} as an effective momentum which, in turn, depends on the mass $M$ via Eq.~\eq{eq:J}. In Fig.~\ref{fig:q:mls}, we plot the solutions $q_{ml}$ for a few values of the angular momentum, $m=0,1,2,3$ as a function of the normalized mass $M R$. As expected, as positive values of mass $M>0$ the solutions $q_{ml}$ are smooth functions of the mass $M$. However, at negative values of $M$ the branches of $q_{ml}$ solutions terminate at certain quantized values of the mass $M$ where these solutions touch the $q=0$ axis. More precisely, the $q_{ml}$ solutions with fixed $l=1$ and variable $m=0,1,\dots$\footnote{Notice that positive and negative $m$ are related by the symmetry~\eq{eq:qml:eq} with respect to the flips of the orbital number $m \to -m - 1$.} terminate at the quantized values of the masses, $MR=-1-m$:
\beqn
q_{m1}(M){\biggl|}_{MR = -1-m} = 0\,.
\label{eq:q:zero}
\eeqn
In fact, the existence of the solutions are proved analytically by expanding the Bessel function around $x\sim 0$. At a slightly lower mass the lowest-$q$ solution becomes $q_{m2}$ thus causing a set of the discontinuities in the lowest-energy eigenvalue~\eq{eq:E:j} that we observe in Fig.~\ref{fig:emum}(d). Notice that the solutions with $l \geqslant 2$ do not vanish so that the property~\eq{eq:q:zero} is only enjoyed by the lowest $l=1$ modes.

\begin{figure}[!thb]
\begin{center}
\vskip 5mm
\begin{tabular}{ccc}
\hskip 3mm {\footnotesize{$M R = -5$}} & \hskip -3mm {\footnotesize{$M R = 0$}} & \hskip -3mm {\footnotesize{$M R = 5$}} \\[-1mm]
\hskip -1mm\includegraphics[scale=0.2455,clip=true]{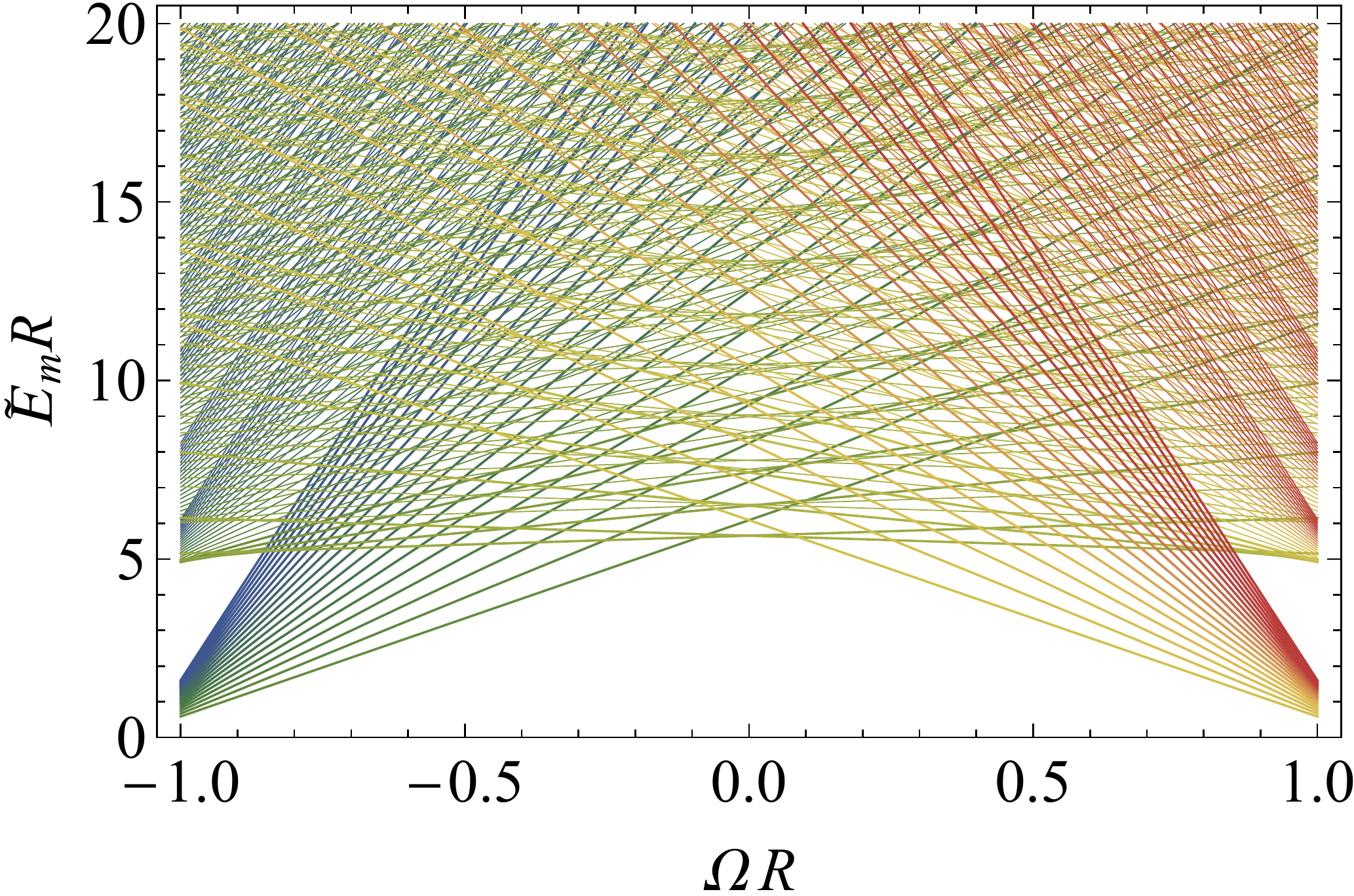} & 
\hskip -3mm \includegraphics[scale=0.225,clip=true]{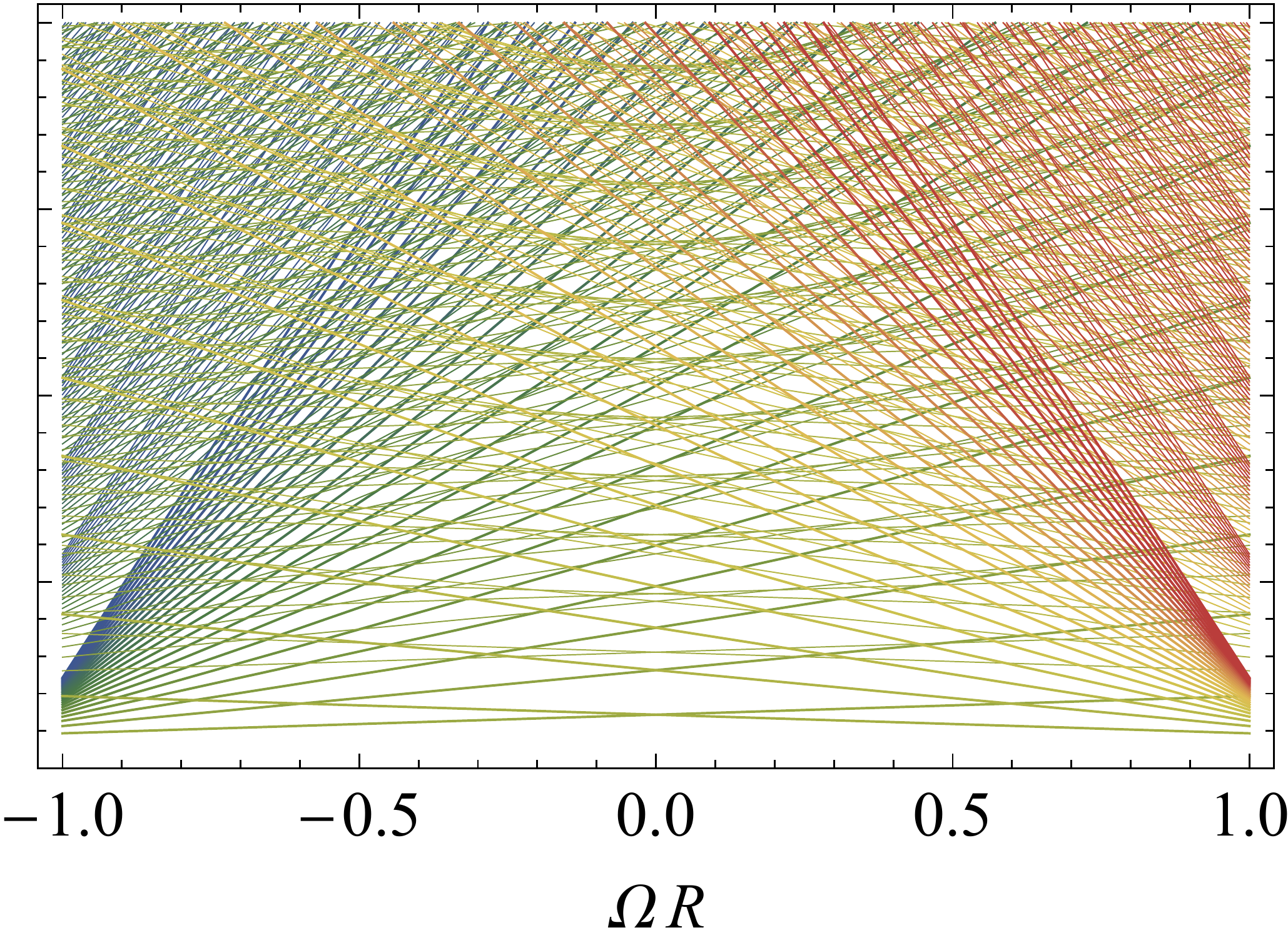} &
\hskip -3mm \includegraphics[scale=0.225,clip=true]{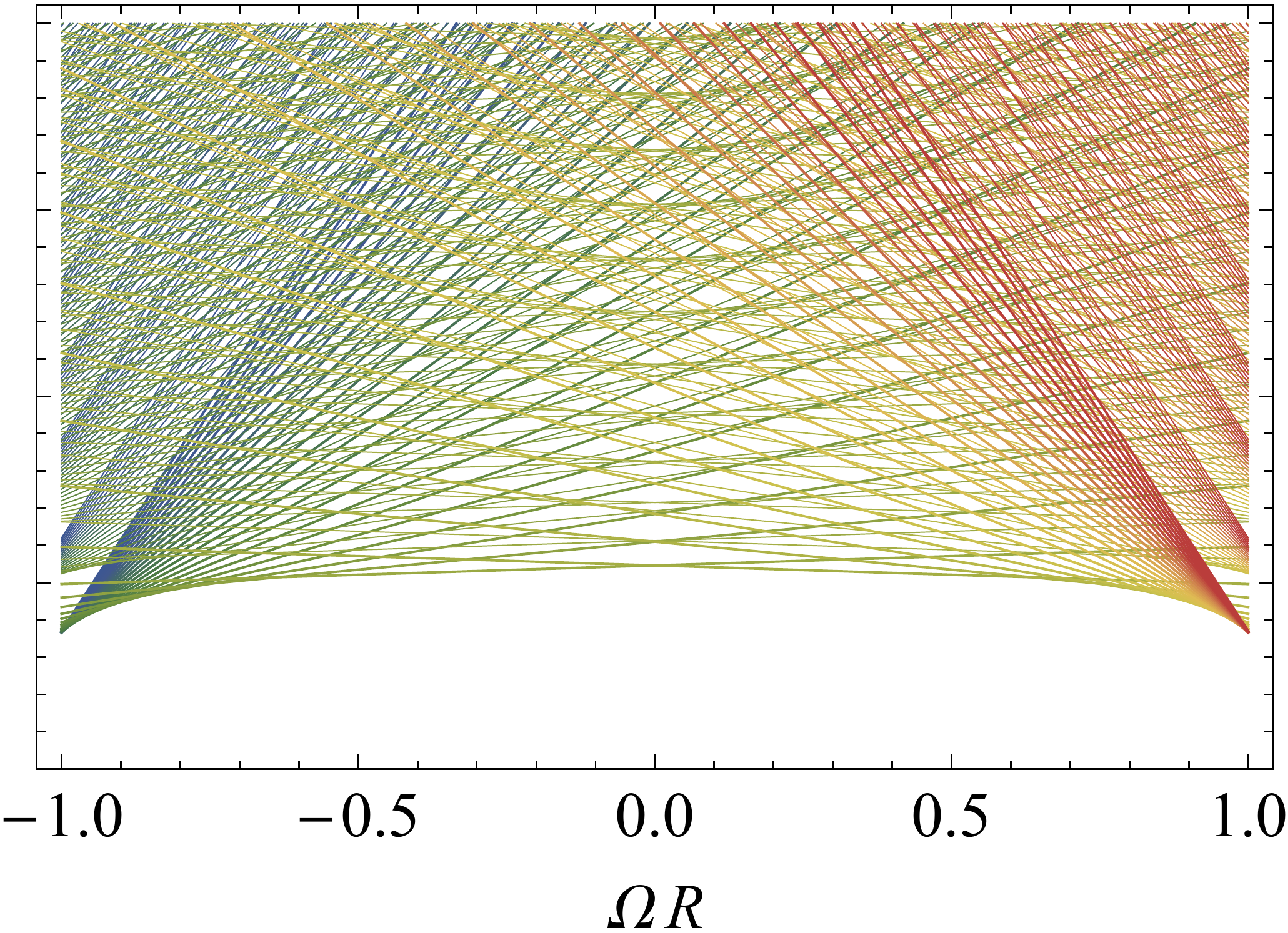}\\
\hskip 3mm (a) & \hskip -3mm (b) & \hskip -3mm (c) 
\end{tabular}
\end{center}
\vskip -5mm
\caption{The energy spectrum in the corotating frame~\eq{eq:Energy} at vanishing longitudinal momentum $k_z = 0$ vs.  the rotation frequency $\Omega$ for (a) positive, (b) zero and (c) negative masses~$M$. The colors vary gradually from blue [the total angular momentum~\eq{eq:mu:m} is negative $\mu_m <0$] via green/yellow ($\mu_m \sim 0$) to red ($\mu_m >0$). The towers in the energy spectrum correspond to the radial excitation number $l=1,2,\dots$. The first tower ($l=1$) is shown by thicker lines.}
\label{fig:spectra}
\end{figure}

Figure~\ref{fig:q:mls} illustrates the asymmetry of the fermionic spectrum in the cylindrical cavity with respect to a change of the sign of the fermion mass $M \to -M$. This asymmetry is a particular feature of the fermionic modes in the cylindric finite-volume geometry with the MIT boundary condition at cylinder's surface~\eq{eq:MIT:boundary}. The MIT boundary condition~\eq{eq:MIT:boundary} breaks down the chiral symmetry explicitly as it is not invariant under the chiral rotations:
\beqn
\psi \to e^{- i \theta \gamma_5} \psi\,, 
\qquad
\bar\psi \to \bar\psi e^{- i \theta \gamma_5}\,.
\label{eq:chiral:transformations}
\eeqn
In the Dirac equation~\eq{eq:Dirac} the chiral transformations~\eq{eq:chiral:transformations} lead to the modification of the mass term: 
\beqn
M \to e^{- i \theta \gamma_5} M e^{- i \theta \gamma_5}\,.
\eeqn
In particular, the choice $\theta = \pi/2$ leads to the flip of the fermion mass, $M\to - M$. Thus, the chirally non-invariant boundary conditions lead to asymmetry of the fermionic spectrum with respect to the flip of the fermion mass. As we will show below, this asymmetry is essential for the dynamical symmetry breaking at small radii of the static cylinder and it becomes less important at larger radii. However, we will also point out that the asymmetry is quite essential for the rotating cylinder. 

Finally, we would like to notice that there exists a ``chiral'' version of the MIT boundary conditions~\eq{eq:MIT:boundary} which is given by a simple flip of a sign in Eq.~\eq{eq:MIT:boundary}:
\beqn
\bigl[i \gamma^\mu n_\mu(\varphi) + 1 \bigr] \psi(t,z,\rho,\varphi) {\biggl |}_{\rho = R} = 0\,. \qquad \qquad \mbox{[chiral MIT b.c.]}
\label{eq:chiral:boundary}
\eeqn
This boundary condition in the cylindrical geometry was considered in Ref.~\cite{Ambrus:2015lfr}.

The MIT boundary condition~\eq{eq:MIT:boundary} may be transformed into the chiral MIT boundary condition~\eq{eq:chiral:boundary} and vice-versa by the chiral transformation~\eq{eq:chiral:transformations} with the chiral angle $\theta = \pi/2$. The same transformation flips the sign of the mass term, $M \to - M$. Therefore all features of the system (i.e. mass spectra, Figs.~\ref{fig:emum} and \ref{fig:spectra}, etc) of the fermions of positive mass subjected to the MIT conditions are the same for the negative-mass fermions satisfying the chiral MIT conditions. In other words, the would-be inequivalence of  features of fermions with positive and negative masses is caused by the boundary conditions which break explicitly the flip symmetry $M \to - M$.

\subsection{Rotation vs external magnetic field}
\label{sec:rotation:vs:magnetic}

In the next section we will make use of the fermionic spectrum~\eq{eq:Energy} to study the phase structure of the interacting fermions in the rotating environment. However, before finishing this Section we would like to make a comment on possibility, in the present relativistic context, of existence of a relation between rotation and (effective) magnetic field which was discussed in the literature recently in Refs.~\cite{Chen:2015hfc,McInnes:2016dwk}.

In the nonrelativistic quantum mechanics the effects of rotation may be represented by an effective uniform magnetic field with the axis parallel to the axis of rotation. The particles in the system then become ``electrically'' charged with respect to the corresponding fictitious ``electromagnetic'' field that represents the rotation. In this Section we show that the effect of a rigid rotation on the spectrum of a relativistic fermion system is generally not equivalent to the effect of the background magnetic field.

In Fig.~\ref{fig:spectra} we show the (co)rotating energy spectrum~\eq{eq:Energy} as a function of the rotation frequency $\Omega$ at vanishing longitudinal momentum $k_z = 0$. From the behavior of the energy spectrum one can figure out at least two arguments against the identification of rotation with external magnetic field:

\begin{enumerate}

\item {\bf Dimensional reduction and energy gap}. \\
{\sl{In a background magnetic field}} $B$ the energy gap between the ground state [the Lowest Landau Level (LLL)] and the first excited level becomes wider as the strength of the magnetic field increases. This leads to the effect of dimensional reduction of particle's motion: in strong magnetic field the particle reside at the LLL as it cannot be excited to the higher energy level due to large energy gap. Since a particle at the LLL may propagate only along the axis of magnetic field, the restriction of particle's energy level to the LLL leads to the dimensional reduction of its physical motion.

{\sl{In a rotating system}}, on the contrary, the energy levels of free fermions do not show dimensional reduction with increasing angular frequency $\Omega$. Indeed, according to Eq.~\eq{eq:Energy} the energy gap between the levels labelled by radial excitation number $l$ does not depend on the angular frequency $\Omega$ at fixed angular momentum $m$. From Fig.~\ref{fig:spectra} we also see that the energy gap between the ground state and the first excited state does not increase. On the contrary, the general tendency is that the energy gap reduces for all fermion's masses $M$ as the rotation frequency increases.  

\item {\bf Ground state degeneracy}.\\ 
{\sl{In a background magnetic field}} $B$ the energy levels are degenerate. The degeneracy factor -- defined as the density of states per unit area of surface transverse to magnetic field --  is a linearly increasing function of the magnetic field strength, $|eB|/(2\pi)$.

{\sl{In a rotating system}}, on the contrary, there is no degeneracy apart from occasional level crossing, Fig.~\ref{fig:spectra}. Moreover, the rotation lists off the double degeneracy of the energy spectrum~\eq{eq:Energy} that emerges between the states with opposite total momentum $\pm \mu_m$ due to relation~\eq{eq:qml:eq}. 

\end{enumerate}

We would like to stress that the rotation leads to the rotational steplike features which are similar to the Shubnikov--de Haas oscillations in magnetic field. This similarity is a direct consequence of the finite-sized geometry and it is not directly linked to the rotation (see Sections~\ref{sec:zero:temperature} and \ref{sec:phase} for more details). It is also important to note that the rotational SdH steps occur in the absence of both external magnetic field and Fermi surface contrary to the conventional magnetic SdH oscillations. 

As the radius $R$ increases the rotational SdH steps become smaller. Figure~\ref{fig:F:sigma:vacuum} suggests that in the limit $R \to \infty$ all discontinuities eventually disappear.

Below we show that at zero-temperature and zero chemical potential the system provides no response to the global rigid rotation, in agreement with Ref.~\cite{Ebihara:2016fwa}. The absence of any $T=0$ response with respect to the rotation comes in sharp contrast with the effect of magnetic field which leads to an enhancement of the chiral symmetry breaking in the vacuum at zero temperature (this effect is known as the magnetic catalysis, Ref.~\cite{Miransky:2015ava}). However, at finite temperature the overall influence of global rotation on the chiral transition is similar to the one of the inverse magnetic catalysis~\cite{Miransky:2015ava}: the critical temperature of the chiral transition is a decreasing function of both temperature and angular frequency~\cite{McInnes:2016dwk}\footnote{A relevant phase diagram at finite radius of the cylinder is plotted below in Fig.~\ref{fig:phase:T}. It has also been discussed in unbounded space in Ref.~\cite{Jiang:2016wvv}.}.  A qualitatively similar effect on the critical temperature may also be produced by a finite chemical potential~\cite{Chen:2015hfc}, so that the analogy of the rotation and magnetic field at finite temperature is not clear.

Thus, we conclude that the energy spectrum shows that the rotation in the relativistic fermion system cannot identically be associated with an effective fictitious magnetic field contrary to the rotation of nonrelativistic fermions as there are strong qualitative differences in responses of the fermionic system with respect to the rotation and magnetic field.

\section{Interacting fermions in rotating frame}
\label{sec:interacting}

A simplest description of interacting fermions is given by the Nambu--Jona-Lasinio (NJL) model. In the rotating frame the NJL Lagrangian for the single fermion species of the mass $m_0$ is given by the following formula:
\begin{align}
S_{\mathrm{NJL}} &= \int_{\mathrm{V}} d^4x \sqrt{- \mathrm{det}\left(g_{\mu \nu}\right)} \, \cL_{\mathrm{NJL}}\left(\bar{\psi},\psi\right), \notag \\
\cL_{\mathrm{NJL}} & = \bar{\psi} \left[ i\gamma _\mu\left(\partial^\mu + \Gamma ^\mu \right) - m_0 \right]\psi +\frac{G}{2}\left[\left(\bar{\psi}\psi\right)^2+ \left(\bar{\psi}i\gamma_5 \psi \right)^2 \right]\,,
\label{eq:L:NJL}
\end{align}
where $g_{\mu\nu}$ corresponds to the metric~\eq{eq:metric} in the rotating reference frame~\eq{eq:varphi} and $\Gamma^\mu$ is the connection~\eq{eq:Gamma}.

Similarly to the case of free fermions, we consider the infinitely long cylinder of radius~$R$, Fig.~\ref{fig:cylinder}. For self-consistency the radius of the cylinder should lie within the light cylinder, $R\Omega \le 1$, so that the rotational velocity at the boundary should not exceed the speed of light. At the cylinder, the fermion modes satisfy the MIT boundary conditions~\eq{eq:MIT:boundary}. The system rotates as a rigid body so that the rotation can be described by a single frequency $\Omega$. Below we assume that the phase is uniform, so that the ground state characteristics (for example, the quark condensate) are coordinate-independent quantities.

The partition function of the NJL model~\eq{eq:L:NJL},
 \begin{align}
\cZ = \int D\psi D\bar{\psi} \exp\left\{i\int_V d^4x \, \cL_{\mathrm{NJL}}\right\}\,,
\label{eq:Z:NJL:1}
\end{align}
can be partially bosonized by inserting the identity 
\begin{align}
1=\int D\pi D\sigma \exp\left\{- \frac{i}{2G} \int _V d^4x \left[\left(\sigma + G\bar{\psi}{\psi}\right)^2 + \left(\pi + G\bar{\psi}i\gamma _5{\psi}\right)^2\right] \right\}\,,
\label{eq:identity}
\end{align}
into Eq.~\eq{eq:Z:NJL:1} and performing the integral over the fermionic fields with the following result:
\begin{align}
\cZ= \int D\pi D\sigma \exp\left\{- \frac{i}{2G} \int_V d^4 x \left(\sigma^2 + \pi^2\right) + \, \mathrm{ln}\,\mathrm{Det} \, \bigl[ i\gamma _\mu\left(\partial^\mu + \Gamma ^\mu \right) - m_0 - \left(\sigma + i\gamma _5\pi \right)\bigr] \right\}.
\label{eq:Z:NJL:2}
\end{align}
In order to study effects of rotation on the dynamical symmetry breaking we put the fermionic current mass -- which breaks the chiral symmetry explicitly -- to zero, $m_0 = 0$. Moreover, we notice that we can always chirally rotate, with a suitably chosen chiral angle~$\theta$, the expression under the fermionic determinant in the partition function~\eq{eq:Z:NJL:2}:
\beqn
& & \mathrm{Det}\left[ i\gamma _\mu\left(\partial^\mu + \Gamma ^\mu \right) - \left(\sigma + i\gamma _5\pi \right)\right] 
 = \mathrm{Det} \left[e^{-i\gamma _5 \theta} \left\{ i\gamma _\mu \left(\partial^\mu + \Gamma ^\mu \right) - \left(\sigma + i\gamma _5\pi \right) \right\}e^{-i\gamma _5 \theta}\right] \notag \\
&= & \mathrm{Det} \left[i\gamma _\mu \left(\partial^\mu + \Gamma ^\mu \right) - e^{-i\gamma _5 \theta}  \left(\sigma + i\gamma _5\pi \right) e^{-i\gamma _5 \theta}\right] =  \mathrm{Det} \left[  i\gamma _\mu \left(\partial^\mu + \Gamma ^\mu \right) - \tilde\sigma \right],
\label{eq:chain:Det:1}
\eeqn
where we applied the chiral transformations~\eq{eq:chiral:transformations} and used the fact that $\mathrm{det}\left(e^{-i\gamma _5 \theta} \right) = 1$. Next, we rotate the scalar-pseudoscalar combination
\beqn
\sigma + i\gamma _5\pi \to e^{-i\gamma _5 \theta} \left(\sigma + i\gamma _5\pi \right) e^{-i\gamma _5 \theta} = \tilde \sigma (\sigma,\pi),
\label{eq:sigma:pi}
\eeqn
into the purely scalar direction determined by new scalar field $\tilde\sigma$ which has an arbitrary sign, $\tilde \sigma = \pm |\tilde \sigma|$, and the absolute value $|\tilde \sigma| = \sqrt{\sigma^2 + \pi^2}$. 

We stress that the sign of the field $\tilde \sigma$ in Eq.~\eq{eq:sigma:pi} should be kept arbitrary because the field $\tilde \sigma$ plays the role of the fermionic mass as it is seen from the expression for the fermion determinant ~\eq{eq:chain:Det:1}. Indeed, the structure of the energy spectrum of free rotating fermions depends -- as we mentioned in Section~\ref{sec:structure} -- on the sign of the fermion mass. For the system of interacting fermions the mass (i.e. its absolute value {\emph{and}} sign) will be chosen dynamically. As we will see below, in the chirally symmetric phase the system would unexpectedly prefer to choose the mass with a negative sign due to the MIT boundary conditions~\eq{eq:MIT:boundary}. 

The measure of integration in the partition function~\eq{eq:Z:NJL:2} is invariant under the transformation~\eq{eq:sigma:pi}. Since the rotation does not couple specifically to the pion mode $\pi$, then in Eq.~\eq{eq:Z:NJL:2} we may always rotate the combination of the pion fields $\sigma$ and $\pi$ to the $\sigma$ direction using the chiral rotation~\eq{eq:sigma:pi}. In the mean-field approximation we neglect fluctuations of the pion fields, so that we may always set $\pi = 0$ and ascribe the effects of the mass gap generation to the constant, coordinate-independent mean-field field $\sigma$. This homogeneous approximation should work for small values of the rotation frequency $\Omega$.

The density of the Helmholtz free energy (the thermodynamic potential) of the system in the (co)rotating frame can be read off from the partition function~\eq{eq:Z:NJL:2}:
\beqn
{\widetilde F}(\sigma,\pi)  & = & \frac{\sigma^2}{2G} + V\bigl(\sigma\bigr)\,,
\label{eq:W:1}
\eeqn
where  
\beqn
V (\sigma) & = & - \frac{i}{{\mathrm{Vol}_4}} \, \mathrm{ln}\, \mathrm{Det} \left[  i\gamma _\mu \left(\partial^\mu + \Gamma ^\mu \right) - {\sigma}\right]
\equiv - \frac{i}{{\mathrm{Vol}_4}} \, \mathrm{tr}\,\mathrm{ln} \, \Bigl[\bigl(i\partial_t+\Omega {\hat J}_z\bigr)^2 + {{\vec \partial}}^{\,2} - {\sigma} ^2\Bigr],
\qquad\quad
\label{eq:V:1}
\eeqn
is the potential induced by the vacuum fermion loop and ${\mathrm{Vol}_4} = \int_V d^4x$ is the volume of the (3+1) dimensional space-time. The fermionic determinant in Eq.~\eq{eq:V:1} has been rewritten using the following chain of identities:
\beqn
& & \mathrm{Det} \left[  i\gamma _\mu \left(\partial^\mu + \Gamma ^\mu \right) - {\sigma}\right]
=\left(\mathrm{Det} \left[  i\gamma _\mu \left(\partial^\mu + \Gamma ^\mu \right) - {\sigma}\right]\mathrm{Det} \left[  i\gamma _\mu \left(\partial^\mu + \Gamma ^\mu \right) - {\sigma}\right] \right)^{\frac{1}{2}}\notag \\
& = & \left(\mathrm{Det} \left[  i\gamma _\mu \left(\partial^\mu + \Gamma ^\mu \right) - {\sigma}\right]\mathrm{Det}\gamma_5 \left[  i\gamma _\mu \left(\partial^\mu + \Gamma ^\mu \right) - {\sigma}\gamma_5 \right]\right)^{\frac{1}{2}} \notag \\
& = & \left(\mathrm{Det} \left[  i\gamma _\mu \left(\partial^\mu + \Gamma ^\mu \right) - {\sigma}\right]\mathrm{Det} \left[i\gamma _\mu \left(\partial^\mu + \Gamma ^\mu \right) + {\sigma}\right] \right)^{\frac{1}{2}} \nonumber \\
& = & \Bigl(\mathrm{Det}\Bigl[\bigl(i\partial_t+\Omega {\hat J}_z\bigr)^2 + {{\vec \partial}}^{\,2} - {\sigma} ^2\Bigr]\Bigr)^{\frac{1}{2}},
\eeqn
where ${{\vec \partial}}^{\,2}$ is the spatial Laplacian. We also used the identity $\mathrm{ln} \, \mathrm{det} \equiv \mathrm{tr}\,\mathrm{ln}$ in Eq.~\eq{eq:V:1}.

In an unbounded (3+1) dimensional space the trace of the logarithm in Eq.~\eq{eq:V:1} can generally be represented as follows:
\beqn
\mathrm{tr} \, \mathrm{ln} \, {\hat {\cal O}} = \int d t \int d^3 x \int \frac{d k_0}{2 \pi} \int \frac{d^3 k}{(2 \pi)^3}\, \mathrm{ln} \, {\cal O}_{k_0,{\bs k}},
\label{eq:integration:unbounded}
\eeqn
where ${\cal O}_{k_0,{\bs k}}$ are the eigenvalues of the operator ${\hat {\cal O}}$. However, in the space bounded by the cylindrical surface the integration over the momentum subspace of the phase space~\eq{eq:integration:unbounded} is modified according to Eq.~\eq{eq:phase:space:k}:
\beqn
\mathrm{tr} \, \mathrm{ln} \, {\hat {\cal O}} = \int d t \int d z \int \frac{d k_0}{2 \pi} \int \frac{d k_z}{2 \pi}\, 
\sum_{l=1}^\infty \sum_{m= -\infty}^\infty  \mathrm{ln} \, {\cal O}_{k_0, k_z, l, m}.
\label{eq:tr:ln}
\eeqn
Consequently, we get for the potential~\eq{eq:V:1} the following expression:
\beqn
V(\sigma) & = & - \frac{i}{\pi R^2} \int \frac{d k_0}{2 \pi} \int \frac{d k_z}{2 \pi}\, \sum\limits_{l=1}^\infty \sum\limits_{m= -\infty}^\infty  \nonumber \\
& & \mathrm{ln} \left[-\left(k_0+\Omega \left(m+\frac{1}{2}\right)\right)^2+k_z^2+\frac{q_{m,l}^2(\sigma)}{R^2} +{\sigma}^2\right].
\label{eq:V:3}
\eeqn

We see that the effect of rotation is to introduce an effective chemical potential $\Omega \mu_m$ for each rotational mode labeled by the rotational quantum number~$m$. The effective chemical potential is the product of the total angular momentum $\mu_m$, Eq.~\eq{eq:mu:m}, and the angular frequency~$\Omega$. Although the term $\Omega \mu_m$ is identified with ``the chemical potential", it is not ab initio clear that the effect of the potential on the rotation is the same as the effect of the standard chemical potential because of the summation over $m$. Notice also that in Eq.~\eq{eq:V:3} the effect of rotation cannot be nullified by a simple shift $k_0 \to k_0 - \mu_m$ for each rotational mode $m$ since the integration over the momentum $k_0$ is carried out over an infinite interval. 

It is also important to mention that the effective potential~\eq{eq:V:3} is generally not symmetric under the reflection $\sigma \to - \sigma$. Indeed, the condensate $\sigma$ enters the potential~\eq{eq:V:3} explicitly, via $\sigma^2$ term, and implicitly, via the eigenvalues $q_{m,l}$ given by the solution of Eq.~\eq{eq:jj}. The condensate $\sigma$ plays a role of the (dynamically generated) mass of the fermion, $M \equiv \sigma$, which affects the spectrum of the eigenvalues $q_{m,l}$  in a nontrivial way, Fig.~\ref{fig:q:mls}. Thus, in general, $q_{m,l}^2(\sigma) \neq q_{m,l}^2(-\sigma)$ and, consequently, $V(\sigma) \neq V(- \sigma)$.

The system at finite temperature is obtained by performing the Wick rotation in Eq.~\eq{eq:V:3}:
\beqn
k_0 \to i \omega_n\,, \qquad \int \frac{d k_0}{2 \pi} \to i T \sum_{n \in \Z}\,, \qquad \omega_n = \pi T \left( n + \frac{1}{2} \right),
\label{eq:omega:n}
\eeqn
where $\omega_n$ is the Matsubara frequency for fermions. Then we get in Eq.~\eq{eq:V:3}:
\beqn
V(\sigma) = - \frac{T}{\pi R^2}  \sum_{m \in \Z} \sum_{n \in \Z} \sum\limits_{l=1}^\infty \int \frac{d k_z}{2 \pi}\, 
\mathrm{ln} \frac{\left(\omega_n - i \Omega \mu_m \right)^2+ E^2_{ml}(k_z,\sigma)}{T^2},
\label{eq:V:4}
\eeqn
where $\mu_m = m + 1/2$ according to Eq.~\eq{eq:mu:m} and the energy $E_{ml}(k_z,\sigma)$ in the laboratory frame is given in Eq.~\eq{eq:E:j}.

Next, we take into account that the sum over the fermionic Matsubara frequencies~\eq{eq:omega:n} using the standard formula:
\beqn
& & \sum_{n \in \Z} \mathrm{ln} \frac{\left(\omega_n - i \Omega \mu_m \right)^2 + E_{ml}^2}{T^2}  =
\frac{1}{2} \sum_{n \in \Z} \left( \mathrm{ln} \frac{\omega_n^2 + \left(E_{ml} - \Omega \mu_m \right)^2}{T^2} 
+ \mathrm{ln} \frac{\omega_n^2 + \left(E_{ml} + \Omega \mu_m \right)^2}{T^2} \right) \nonumber \\
& & \qquad =  \frac{\varepsilon}{T} + \ln \left( 1 + e^{- (E_{ml} - \Omega \mu_m)/T} \right)+ \ln \left( 1 + e^{- (E_{ml} + \Omega \mu_m)/T} \right) + {\mathrm{const}}\,,
\label{eq:sum:1}
\eeqn
which highlights the interpretation of the product of the total angular momentum~\eq{eq:mu:m} and the angular frequency~$\Omega$ as a chemical potential $\Omega \mu_m$ for each eigenmode $m$.

We substitute Eq.~\eq{eq:sum:1} into Eq.~\eq{eq:V:4} and disregard in Eq.~\eq{eq:sum:1} the last term which does not include the condensate $\sigma$ and the angular frequency $\Omega$. The resulting effective potential has two parts:
\beqn
V(\sigma; T, \Omega) & = & V_{\mathrm{vac}}(\sigma) + V_{\mathrm{rot}}(\sigma; T, \Omega)\,,
\label{eq:V:5}
\eeqn
where the first term is the divergent vacuum energy which depends neither on temperature nor on rotation velocity:
\beqn
V_{\mathrm{vac}}(\sigma) = - \frac{1}{\pi R^2} \sum_{m \in \Z} \sum\limits_{l=1}^\infty \int \frac{d k_z}{2 \pi}\, E_{ml}(k_z,\sigma) \,.
\label{eq:V:vac}
\eeqn
The second term in Eq.~\eq{eq:V:5} captures the rotational and temperature effects:
\beqn
V_{\mathrm{rot}}(\sigma;T,\Omega) & = & - \frac{T}{\pi R^2} \sum_{m \in \Z}  \sum\limits_{l=1}^\infty \int \frac{d k_z}{2 \pi}\, \nonumber \\
& & 
\biggl[\ln \left( 1 + e^{- \frac{E_{ml}(k_z,\sigma) - \Omega \mu_m}{T}} \right) + \ln \left( 1 + e^{-\frac{E_{ml}(k_z,\sigma) +\Omega \mu_m}{T}} \right) \biggr]. \qquad
\label{eq:V:rot}
\label{eq:V:thermal}
\eeqn

Generally, the effect of rotation on the particle properties may be understood from the fact that at finite temperature the thermal occupation number of fermionic particles is determined by the energy of the particles calculated in the {\emph{rotating}} frame~\eq{eq:E:j},
\beqn
n(T,\Omega) = \left(1 + e^{- \frac{\tE}{T}} \right)^{-1} \equiv \left(1 + e^{- \frac{E - \Omega \mu_m}{T}} \right)^{-1}\,,
\label{eq:n}
\eeqn
and not by the energy defined in the laboratory frame~\eq{eq:Energy}.

The vacuum part of the potential~\eq{eq:V:vac} is divergent in the ultraviolet limit and therefore it has to be regularized:
\beqn
V_{\mathrm{vac}}(\sigma) = - \frac{1}{\pi R^2} \sum_{m \in \Z} \sum\limits_{l=1}^\infty \int \frac{d k_z}{2 \pi}\, \
f_\Lambda\left(\sqrt{k_z^2 + \frac{q_{ml}^2}{R^2}}\right) E_{ml}(k_z,\sigma) \,.
\label{eq:V:vac:reg}
\label{eq:V:vacuum}
\eeqn
where $f_\Lambda$ is a cutoff function. For this function one may use the exponential cutoff~\cite{Gorbar:2011ya,Chen:2015hfc}:
\beqn
f^{{\mathrm{exp}}}_\Lambda(\varepsilon) = 
\frac{\sinh(\Lambda/\delta \Lambda)}{\cosh(\varepsilon/\delta \Lambda) + \cosh (\Lambda/\delta \Lambda)}\,,
\label{eq:f:Lambda:2}
\eeqn
where one takes phenomenologically $\delta \Lambda = 0.05 \Lambda$ following Ref.~\cite{Chen:2015hfc}.

Summarizing, the free energy of the gas of rigidly rotating fermions in the cylinder is given by Eq.~\eq{eq:W:1}:
\beqn
{\widetilde F}(\sigma)  & = & \frac{\sigma^2}{2G} + V_{\mathrm{vac}}(\sigma) + V_{\mathrm{rot}}(\sigma;T,\Omega)\,,
\label{eq:F:free:energy}
\eeqn
where the vacuum and (rotational) thermal parts of the total potential are given by Eqs.~\eq{eq:V:vacuum} and \eq{eq:V:thermal}, respectively.

We stress that the free energy~\eq{eq:F:free:energy} is defined in the (co)rotating frame which rotates with the angular velocity that matches exactly the angular velocity of rotating fermion medium. One can see that the free energy~\eq{eq:F:free:energy} is determined with respect to the rotating frame since the thermodynamic potential~\eq{eq:V:rot} involves only the energy levels ${\tE}$ determined in the rotating frame\footnote{A rotation-dependent contribution to the vacuum energy~\eq{eq:V:vacuum} cancels out exactly.}. Moreover, the densities of the particles that are associated with the free energy~\eq{eq:F:free:energy} have the Fermi distribution in the rotating frame, $n = (e^{{\tE}/T}+1)^{-1}$. We discuss the (free) energies in the laboratory and (co)rotating frames in Section~\ref{sec:angular}.

\section{Chiral symmetry breaking in a cylinder at zero temperature}
\label{sec:zero:temperature}

\subsection{Zero temperature phase diagram}
\label{sec:T:0:phase}

\begin{figure}[!thb]
\begin{center}
\begin{tabular}{cc}
\includegraphics[scale=0.51,clip=true]{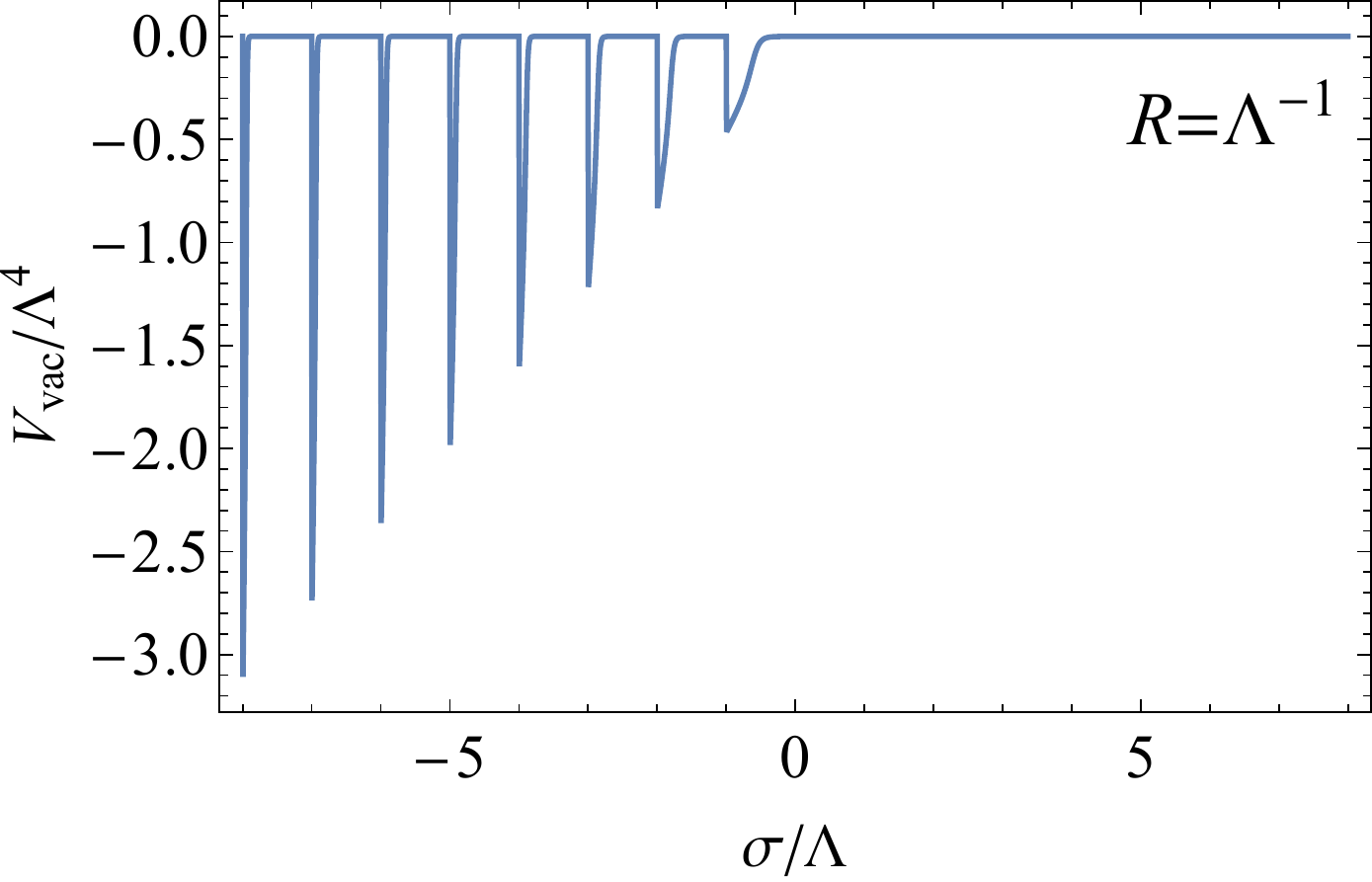} &  \includegraphics[scale=0.52,clip=true]{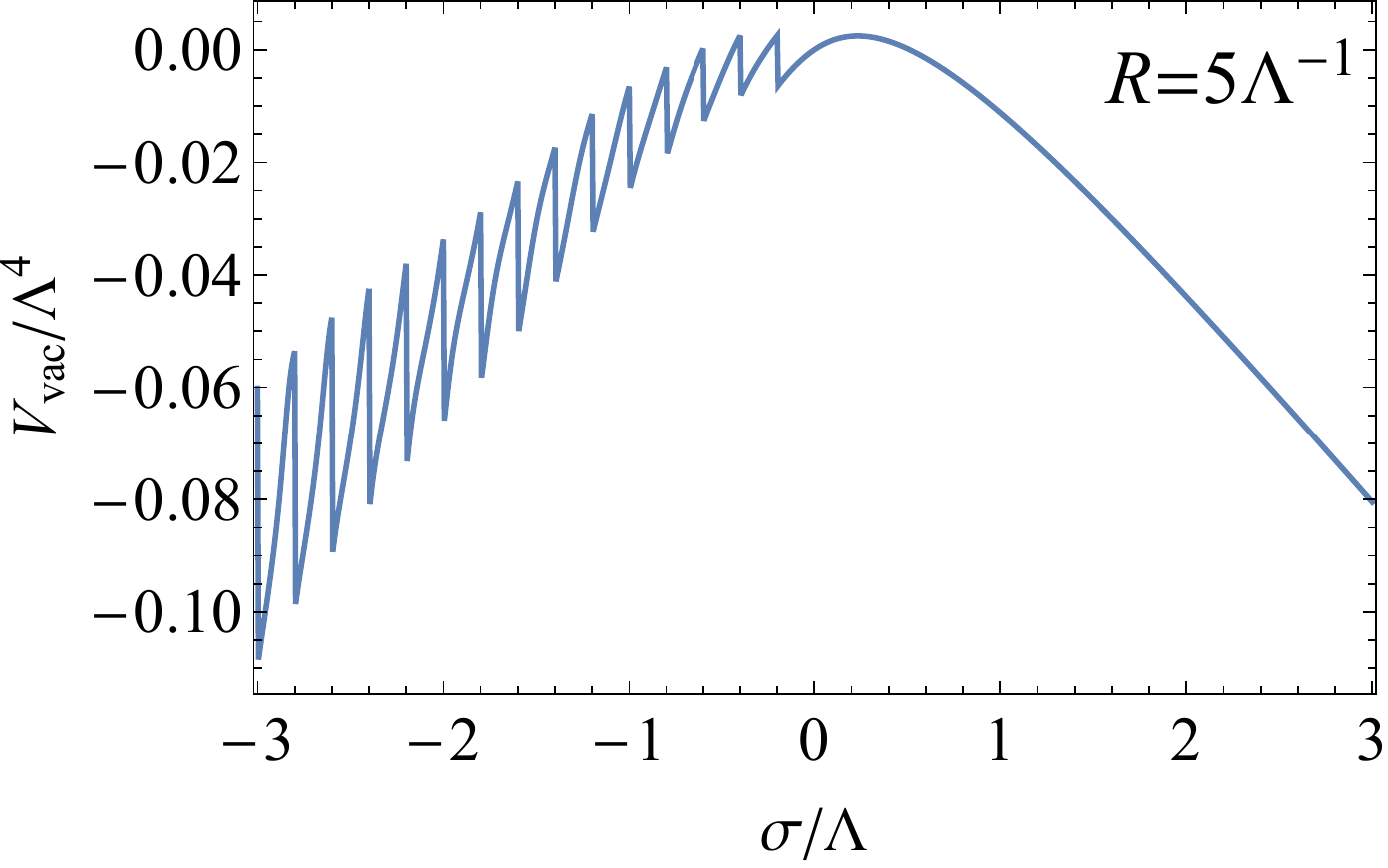} \\
(a) & (b) \\[5mm]
\multicolumn{2}{c}{\includegraphics[scale=0.65,clip=true]{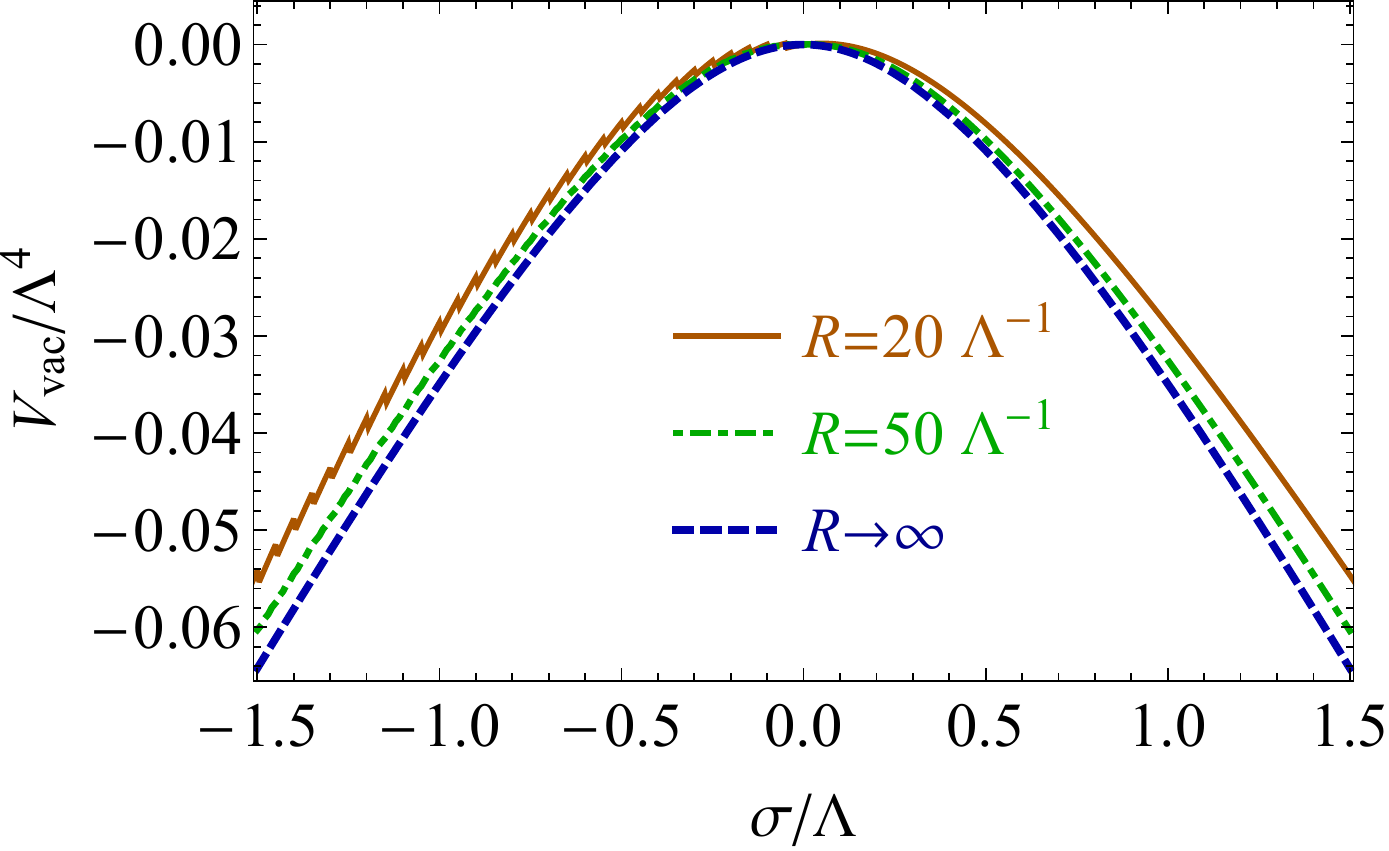}}\\
\multicolumn{2}{c}{(c)}
\end{tabular}
\end{center}
\vskip -3mm
\caption{Vacuum energy~\eq{eq:V:vac:reg} with the cutoff function~\eq{eq:f:Lambda:2} at (a) small, (b) moderate and (c) large values of the cylinder radius $R$. All dimensional quantities are given in terms of the cutoff parameter~$\Lambda$.}
\label{fig:Vvac}
\end{figure}

Let us consider the properties of the vacuum potential~\eq{eq:V:vac:reg} which is regularized with the help of the cutoff function~\eq{eq:f:Lambda:2}. In Fig.~\ref{fig:Vvac}, we show the regularized vacuum potential as the function of the condensate $\sigma$ at different radii of the cylinder~$R$. 

\vskip 3mm

First of all, we notice that at finite radius of the cylinder the vacuum potential is a smoothly diminishing function at all positive values of the condensate $\sigma$. However at negative $\sigma$ the potential exhibits infinite series of local minima which become progressively deeper as the value of the condensate $\sigma$ decreases. One can observe that the potential is a sum of two parts: a smooth potential which is symmetric with respect to the reflection of the potential, $\sigma \to - \sigma$, and a saw-like part at negative values of the condensate $\sigma$. At small radius $R$ the smooth part is very small so that the saw-like part dominates, Fig.~\ref{fig:Vvac}(a). As the radius increases, both parts of the potential becomes comparable with each other, Fig.~\ref{fig:Vvac}(b) and, finally, at large radius the potential is dominated by the smooth component. In the limit $R \to \infty$ the potential becomes a smooth function of $\sigma$ and the symmetry with respect to the flips $\sigma \to - \sigma$ is restored, Fig.~\ref{fig:Vvac}(c). 

The unusual behavior of the vacuum potential observed at a finite radius of the cylinder is an indirect consequence of the presence of the regularizing function~\eq{eq:f:Lambda:2} which effectively suppresses the energies $E_{ml}$, Eq.~\eq{eq:E:j}, that are higher than the value of the cutoff parameter $\Lambda$. This procedure works rather well at positive values of the condensate~$\sigma$ because the energy is a regular, increasing function of the condensate. However at negative values of $\sigma$ the energy~\eq{eq:E:j} becomes a non-monotonic function of the condensate~$\sigma$ due to the successive vanishing of the eigenvalues $q_{ml}$ of Eq.~\eq{eq:jj} with increase of the value of the condensate~$\sigma$, Fig.~\ref{fig:q:mls}. We stress that this particular feature is a specific property of the MIT boundary conditions imposed at the cylindrical surface~\eq{eq:MIT:boundary}. If instead of the MIT boundary conditions we would choose its chiral analog~\eq{eq:chiral:boundary} then the vacuum energy in Fig.~\ref{fig:Vvac} would appear in a mirrored form, $\sigma \to - \sigma$, and consequently the nonmonotonic behavior of the potential would be observed at the positive values of the chiral condensate.

The vacuum part~\eq{eq:V:vacuum} of the potential plays an important role in our discussions below as it defines the ground state of the theory together with the thermal part~\eq{eq:V:thermal} that captures the effects of rotation. The regularization of the vacuum potential by a cutoff function is a very standard procedure which is always implemented in the NJL model in the thermodynamic limit. We thoroughly follow this standard procedure in our case when the physical volume of the model is restricted by the cylinder surface in two dimensions but still stays infinite due to unbounded third dimension along the axis of the cylinder. The implementation of the cutoff is an obligatory requirement, which is, basically, associated with the nonrenormalizability of the NJL model. Notice that the saw-like behavior of the vacuum potential does not depend qualitatively on the particular form of the cutoff function which in our case is given by Eq.~\eq{eq:f:Lambda:2}. Therefore we consider this property of the vacuum potential as a physical feature. We hope that it may eventually be clarified in other, renormalizable models. 

\vskip 5mm
\begin{figure}[!thb]
\begin{center}
\includegraphics[scale=0.6,clip=true]{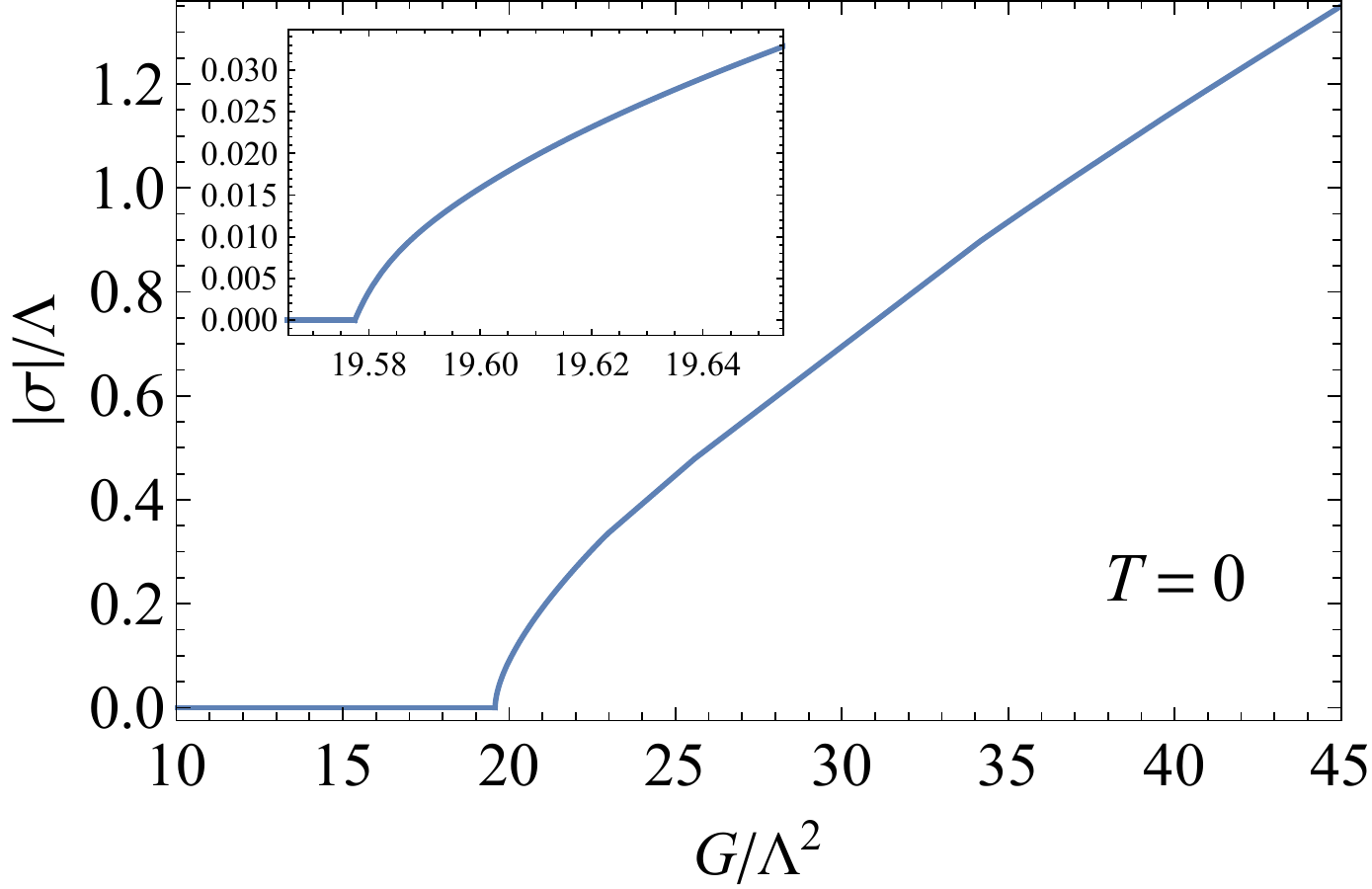} 
\end{center}
\vskip -5mm
\caption{The condensate as a function of the coupling constant $G$ in the ground state in thermodynamic limit $R \to \infty$ at zero temperature. The inset zooms in on the critical region around~$G_s$.}
\label{fig:sigma:thermodynamic}
\end{figure}

In the limit of large radius $R$ the invariance of the potential $V(\sigma)$ on the sign flips $\sigma \to - \sigma$ is restored, as expected. In Fig.~\ref{fig:sigma:thermodynamic} we show the ground-state condensate~$\sigma \equiv \avr{\sigma}$ as a function of the coupling constant $G$. The dynamically broken phase is realized at $G > G_c$, where the critical value of the NJL coupling constant is as follows:
\beqn
G_c \equiv G_c(R \to \infty) = 19.578\, \Lambda^{-2} \,.
\label{eq:Gc}
\eeqn

\begin{figure}[!thb]
\begin{center}
\begin{tabular}{cc}
\includegraphics[scale=0.5, clip=true]{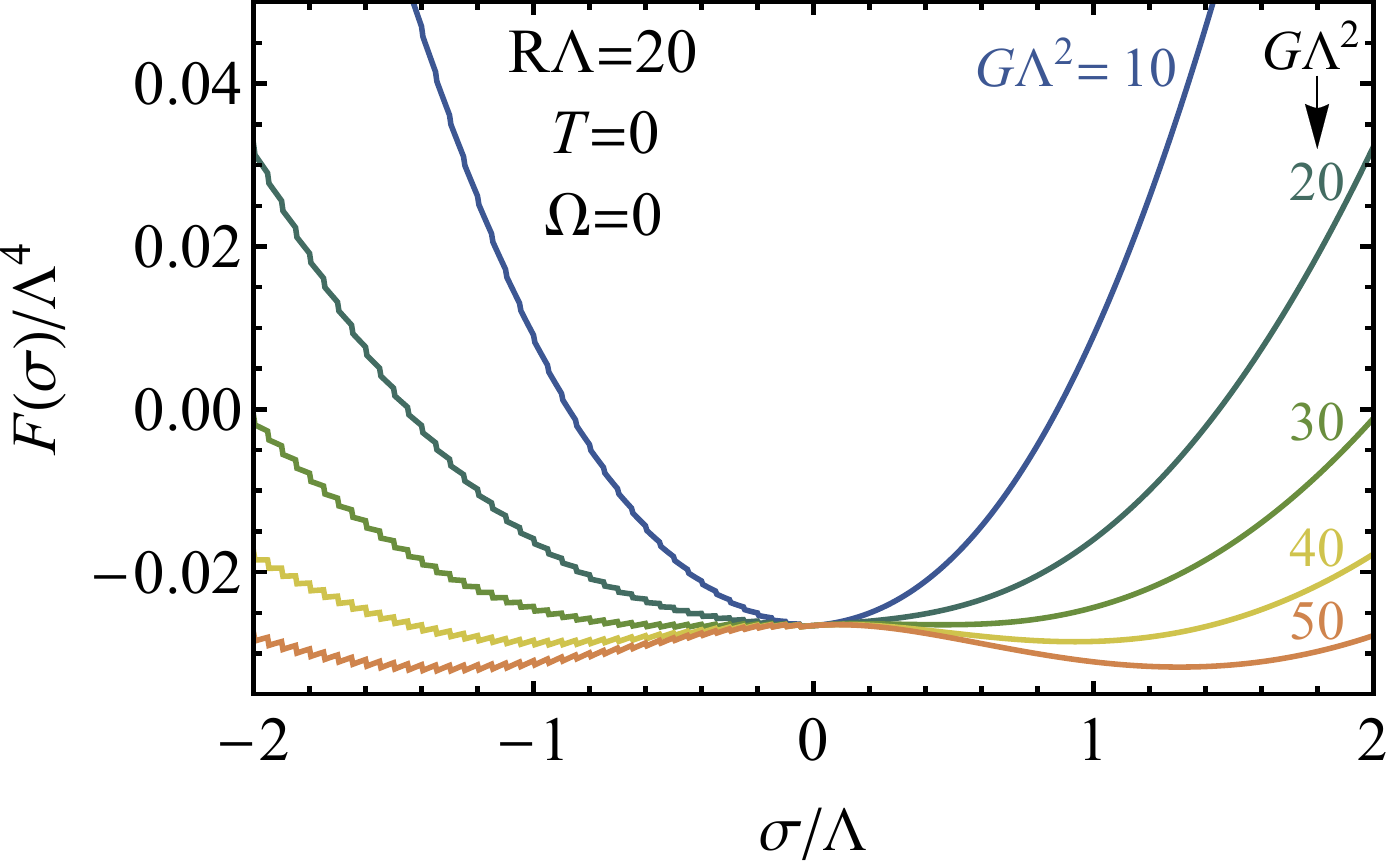} & 
\includegraphics[scale=0.475, clip=true]{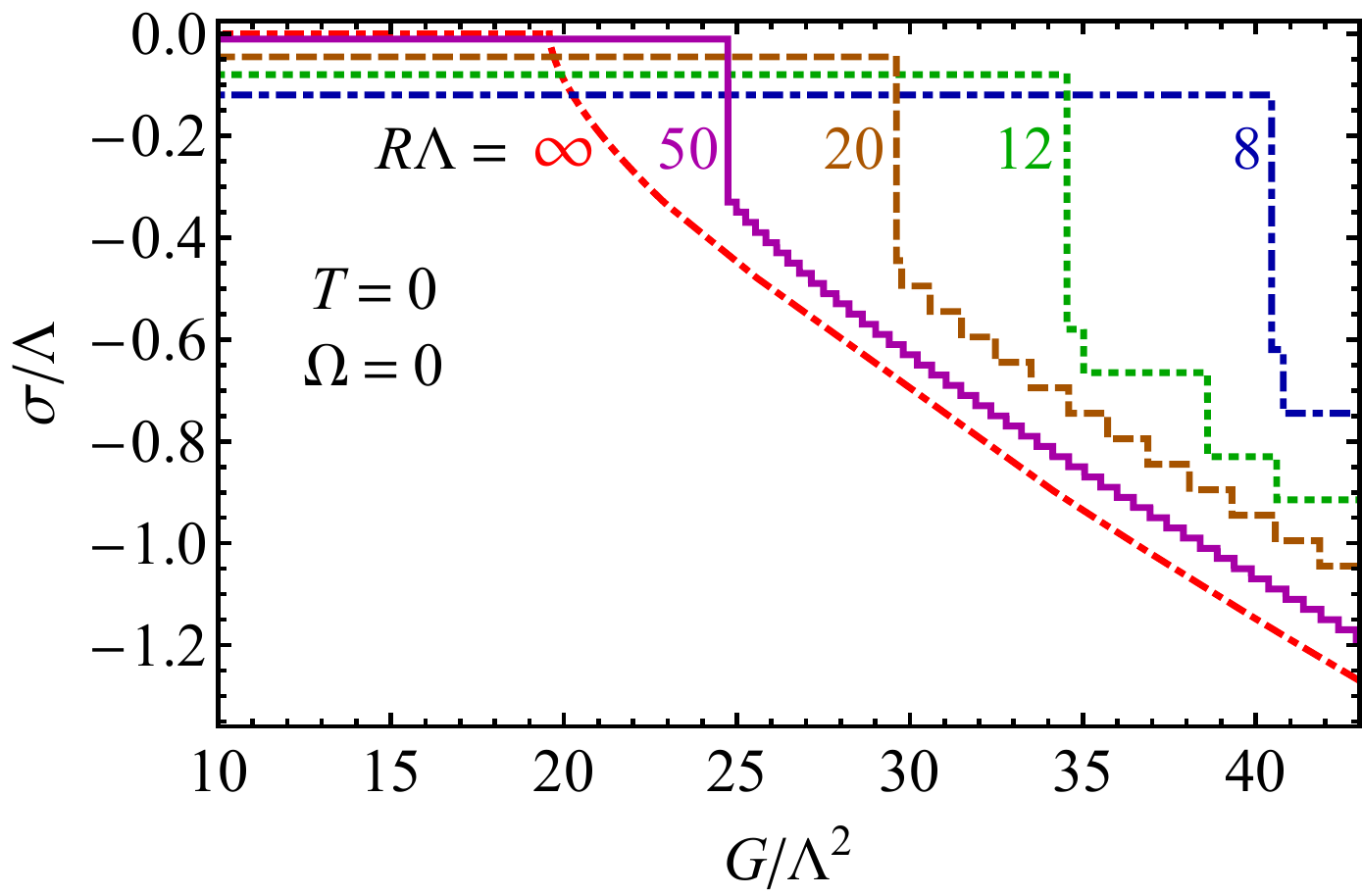} \\
(a) & (b)
\end{tabular}
\end{center}
\caption{(a) Free energy~\eq{eq:F:free:energy} of the nonrotating cylinder at zero temperature as the function of the field $\sigma$ at various values of the coupling constant $G$. (b) The ground-state condensate $\sigma$ as function of the coupling constant $G$ at various fixed radii $R$. The negative branch of the dynamically generated condensate in the infinite volume $R\to \infty$ is also shown for comparison (taken from Fig.~\ref{fig:sigma:thermodynamic}).}
\label{fig:F:sigma:vacuum}
\end{figure}

The free energy~\eq{eq:F:free:energy} of fermions in the nonrotating cylinder of finite radius $R = 20\, \Lambda^{-1}$ at zero temperature is shown in Fig.~\ref{fig:F:sigma:vacuum}(a) at various values of the coupling constant $G$. One notices that the saw-like nature of the potential becomes will-pronounced at the finite radius as negative ($\sigma <0$) and positive ($\sigma > 0$) parts of the potential are visibly different from each other. The direct consequence of this asymmetry is that at finite radius the system prefers to develop a \emph{negative} condensate $\sigma <0$ in the ground state, Fig.~\ref{fig:F:sigma:vacuum}(b). A local minimum at a positive value of $\sigma$ is not a ground state of the model. [Notice that if we would choose the chiral analog of the MIT boundary conditions~\eq{eq:chiral:boundary} then the corresponding free energy would be given by Fig.~\eq{eq:F:free:energy} with the sign flip $\sigma \to - \sigma$ and the system would develop a \emph{positive} condensate $\sigma > 0$ in the ground state.]

The condensation of nonrotating fermions at finite radius $R$ of the cylinder has a few interesting features which are readily seen in Fig.~\ref{fig:F:sigma:vacuum}(b).
\begin{enumerate}

\item 
\underline{Steplike discontinuities in the condensate}. Contrary to the smooth behavior of the condensate in the thermodynamic limit, the condensate $\sigma$ at a finite radius $R$ shows multiple steplike features signaling that the condensate (and, consequently, mass gap) changes discontinuously at (infinite) set of values of the coupling constant $G$.  These predicted steplike features have the same nature as the Shubnikov--de Haas (SdH) oscillations~\cite{ref:SdH} because they occur due to the presence of the discrete energy levels. Although the discreteness of the levels is a natural feature of finite-volume systems, these rotational SdH-like behaviors evolve nontrivially with the angular frequency $\Omega$ [see, for example, Fig.~\ref{fig:phase:T}(a) below] thus suggesting certain similarity of rotation with the presence of external magnetic field (see, however, our discussion in Sect.~\ref{sec:cold:vacuum}). It is crucial to highlight that the rotational SdH steps occur in the absence of both external magnetic field and Fermi surface contrary to conventional SdH oscillations. As the radius $R$ increases the rotational SdH steps become smaller. Figure~\ref{fig:F:sigma:vacuum} suggests that in the limit $R \to \infty$ all discontinuities eventually disappear.

\item \underline{Delayed broken phase}. The critical value $G_c$ of the coupling constant becomes larger as the radius $R$ decreases. In other words, the dynamically broken phase in a cylinder of a finite radius emerges at higher values of $G$ compared to the thermodynamic limit.

\item \underline{Weaker dynamical symmetry breaking}. At finite $R$ the dynamical mass gap generation is smaller compared to the infinite-volume limit. The finite geometry suppresses the condensate $\sigma$.

\item \underline{Explicit symmetry breaking due to finite radius}. There is no dynamical symmetry breaking  small coupling constant $G < G_c$. However, we observe the effect of explicit symmetry breaking: as radius $R$ decreases the condensate $\sigma$ becomes larger. As we will see later this natural effect is associated with the MIT boundary conditions~\eq{eq:MIT:boundary} which violate the chiral symmetry~\eq{eq:chiral:transformations} explicitly.

\end{enumerate}

\begin{figure}[!thb]
\begin{center}
\begin{tabular}{cc}
\includegraphics[scale=0.5, clip=true]{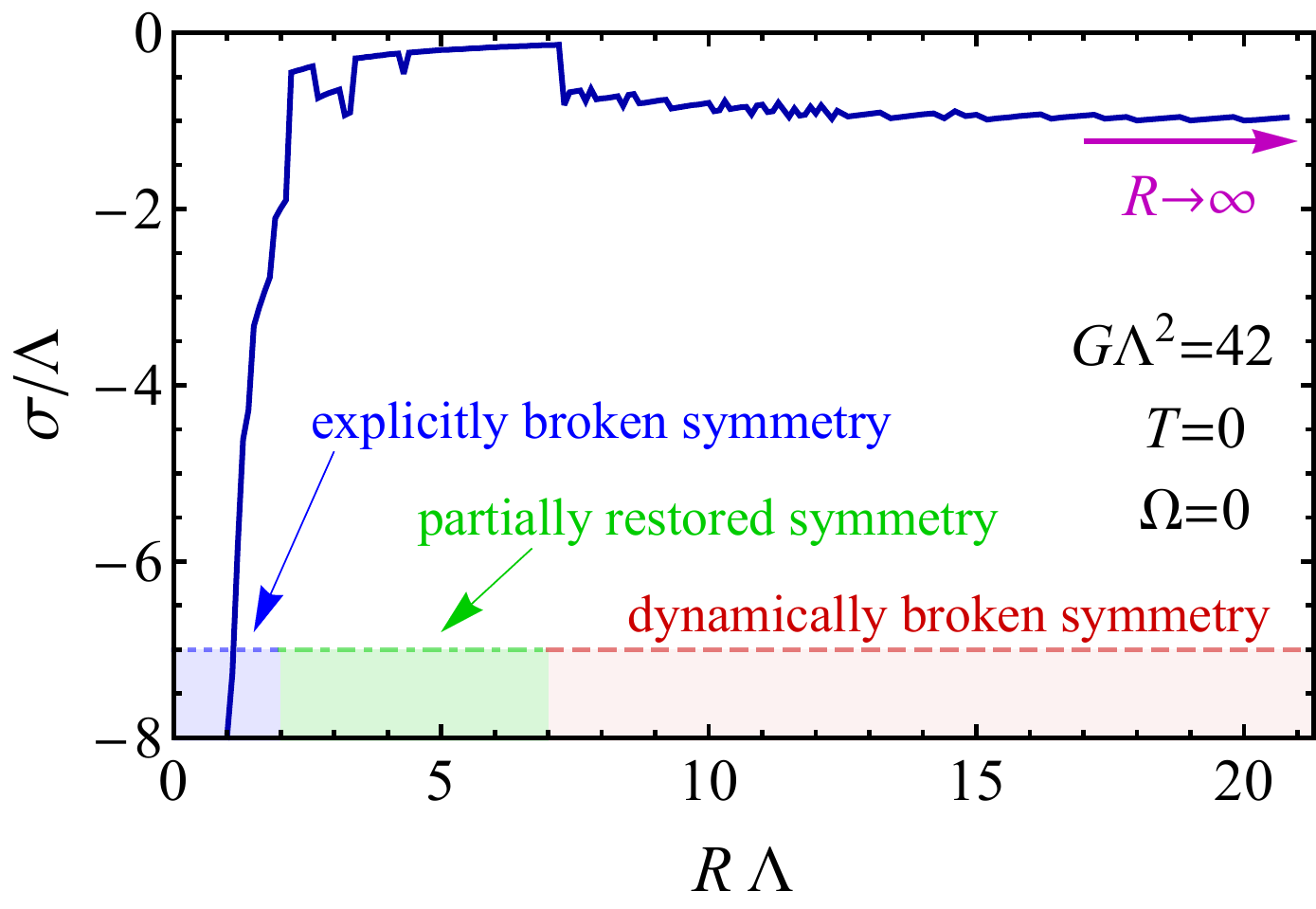} &
\includegraphics[scale=0.505, clip=true]{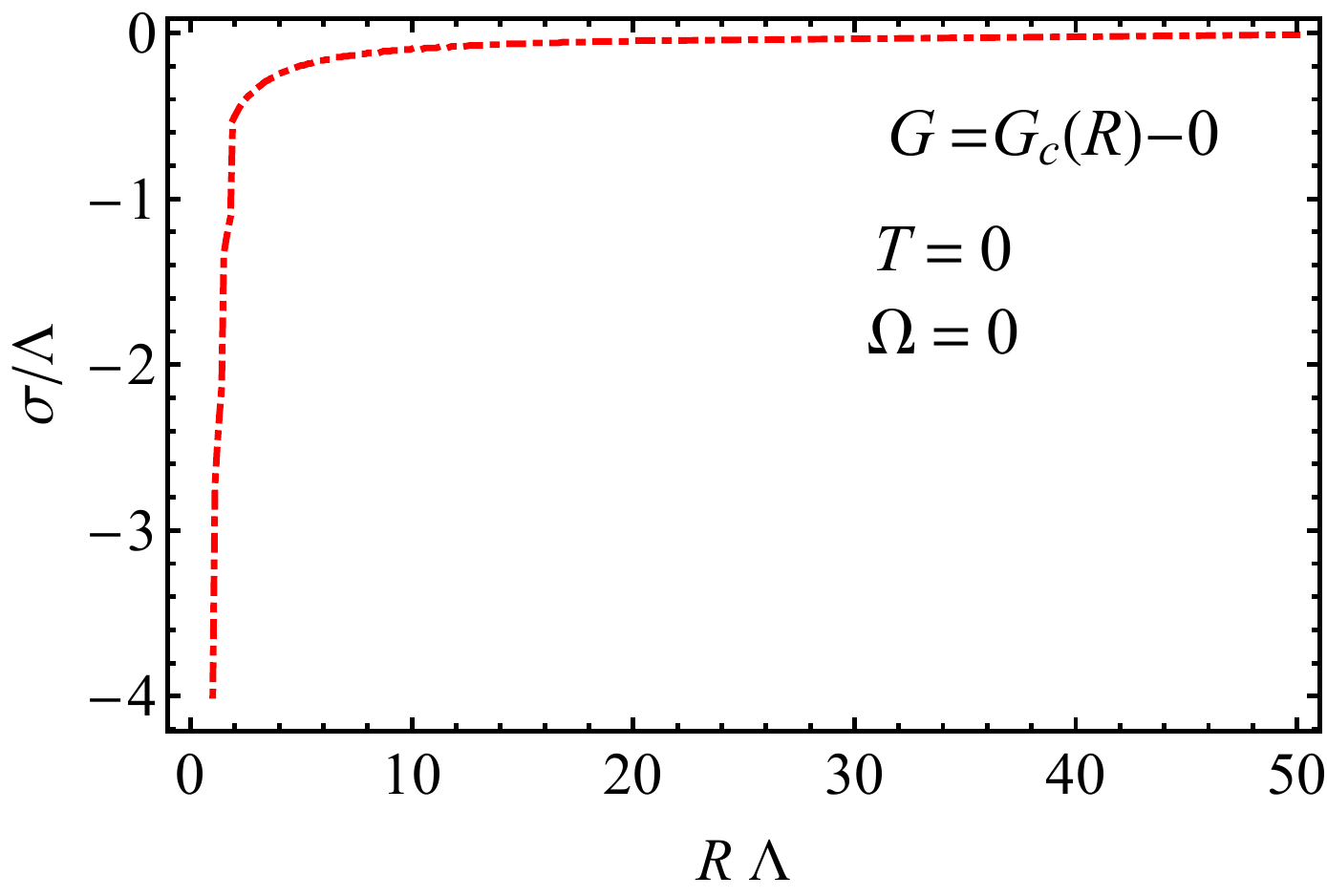} \\
(a) & (b)
\end{tabular}
\end{center}
\caption{(a) Ground-state condensate in a nonrotating cylinder at the coupling $G = 42 \, \Lambda^{-2} > G_c(R \to \infty)$ as the function of the radius $R$ of the cylinder. The value of the condensate in the thermodynamic limit $R \to \infty$ is shown by the arrow. At the bottom of the figure we highlight the explicitly broken, partially restored and dynamically broken phases (regions) which appear, respectively, at small, moderate and large radii of the cylinder. (b) The explicitly induced condensate $\sigma$ which appears in the interior of the cylinder due to  the MIT boundary conditions imposed at its finite radius $R$ at $G = G_c(R) - 0$. At small radii $R\sim\mbox{(a few)} \Lambda^{-1}$ the condensate is large while as the radius of the cylinder increases the explicitly broken chiral symmetry gets restored and the condensate vanishes.}
\label{fig:sigma:T0}
\end{figure}

Thus, in nonrotating cylinder at zero temperature we expect to observe the existence of three different regions in the phase diagram. At small radius of the cylinder the MIT boundary conditions  break the chiral symmetry explicitly by inducing a large nonzero negative value of the condensate $\sigma$ that increases with decrease of the radius $R$. At moderate radii, as the boundary effect becomes weak, the value of the condensate almost disappears and the chiral symmetry is partially restored. At larger radii the chiral symmetry gets broken again, at this case dynamically. The dynamical chiral symmetry breaking at finite cylinder's radius is smaller compared to the thermodynamic limit.  All these features are illustrated in Figs.~\ref{fig:sigma:T0}(a) and (b).

\begin{figure}[!thb]
\begin{center}
\includegraphics[scale=0.55, clip=true]{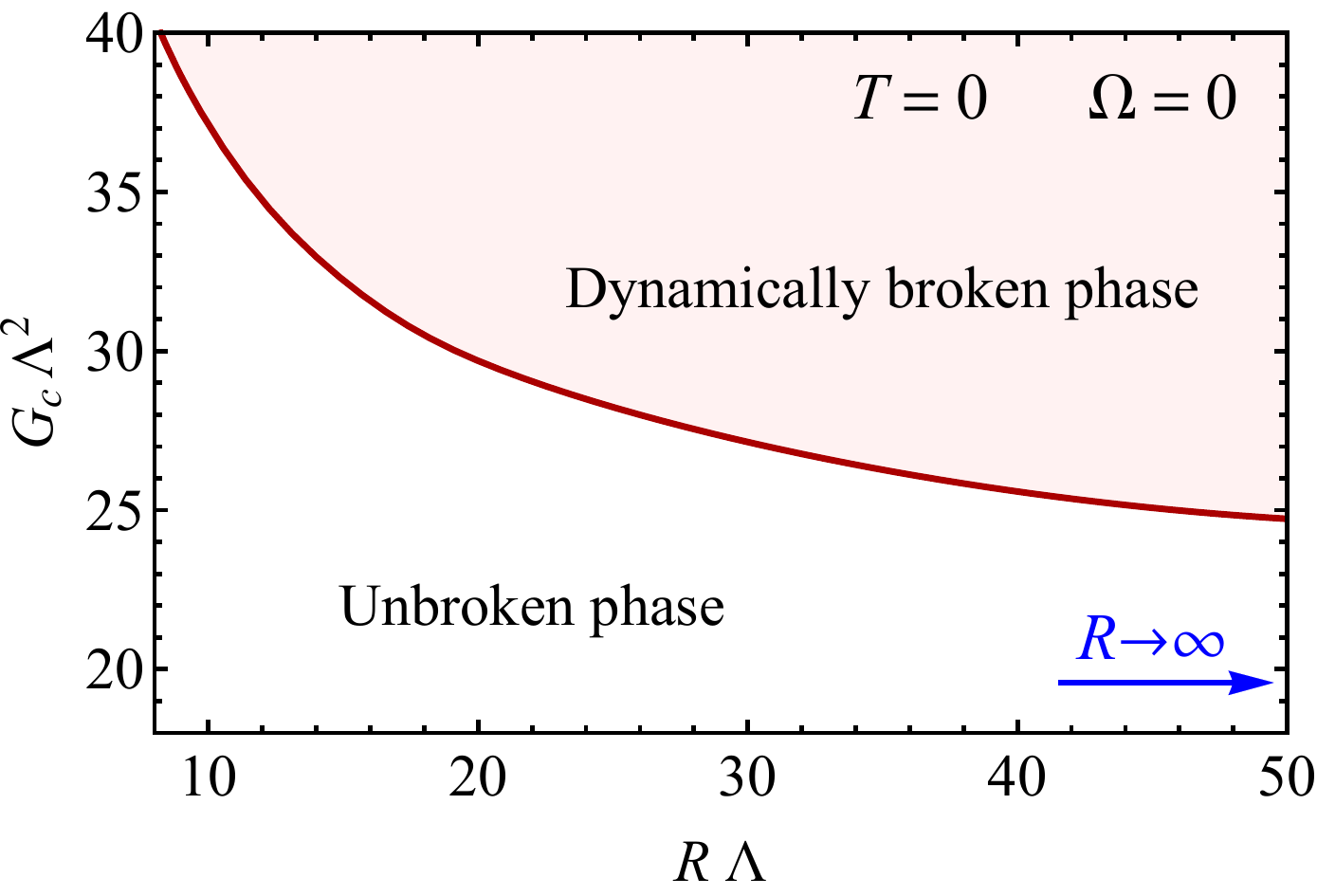} 
\end{center}
\caption{Phase diagram in a cylinder of the finite radius $R$ at zero temperature. We do not show a stretch at small values of the cylinder radius $R\sim\!\mbox{(a few)}/\Lambda$ where the chiral symmetry is explicitly broken by the MIT boundary conditions.}
\label{fig:phase}
\end{figure}

The zero-temperature phase diagram in the $R-G$ plane is shown in Fig.~\ref{fig:phase}. The critical coupling of the chiral transition $G_c(R)$ at finite $R$ is higher than its value in the thermodynamic limit~\eq{eq:Gc}. As the radius $R$ increases the critical coupling $G_c(R)$ becomes smaller. At small radii the chiral symmetry is broken explicitly by the MIT boundary conditions (not shown in Fig.~\ref{fig:phase}).

\subsection{Cold vacuum does not rotate}
\label{sec:cold:vacuum}

At the end of this Section let us consider the response of the rotating fermionic medium in the zero-temperature limit. As we have already seen, the vacuum part~\eq{eq:V:vac:reg} does not depend on the rotating frequency~$\Omega$, and therefore the rotational effects may only be captured by the rotational part of the effective potential~\eq{eq:V:rot}. Using the identity,
\beqn
\lim_{T \to 0} T \ln \left( 1 + e^{- \frac{\varepsilon - \mu}{T}} \right) = (\mu - \varepsilon) \theta(\mu - \varepsilon),
\eeqn
we rewrite the rotational potential~\eq{eq:V:rot} as follows:
\beqn
V_{\mathrm{rot}}^{0}(\sigma;\Omega) \equiv \lim_{T \to 0} V_{\mathrm{rot}}(\sigma;T,\Omega) & = & - \frac{1}{\pi R^2} \sum_{m \in \Z} \sum\limits_{l=1}^\infty \int \frac{d k_z}{2 \pi}\,  (|\mu_m| - E_{ml}) \theta(|\mu_m| - E_{ml}) \nonumber \\
& = & - \frac{2}{\pi R^2} \sum_{m = 0}^\infty \sum\limits_{l=1}^\infty \int \frac{d k_z}{2 \pi}\,  (\mu_m - E_{ml}) \theta(\mu_m - E_{ml}), \qquad
\label{eq:V:rot:2}
\eeqn
where we used the flip symmetry of Eq.~\eq{eq:qml:eq} to simplify the sum over angular momentum variable $m$. An explicit integration gives:
\beqn
V_{\mathrm{rot}}^0  = - 2 \sum_{m = 0}^\infty \sum\limits_{l=1}^\infty \frac{q_{ml}^2 + \sigma^2 R^2}{2 \pi^2 R^4} \left[ \nu_{ml} \sqrt{\nu_{ml}^2 - 1}  
+ \ln \left( \nu_{ml} + \sqrt{\nu_{ml}^2 - 1} \right)\right] \theta(\nu_{ml} - 1),
\qquad 
\label{eq:V:rot:3}
\eeqn
where
\beqn
\nu_{ml} = \frac{(m+1/2) \Omega R}{\sqrt{q_{ml}^2 + { \sigma}^2 R^2}}\,.
\label{eq:nu:ml}
\eeqn

Notice that due to the step function, the sum over the integer variable $l$ in Eq.~\eq{eq:V:rot:3} is always finite. The quantity~\eq{eq:nu:ml} can be expressed via energies $\tE_{ml}$ and $E_{ml}$ in, respectively, the rotating frame~\eq{eq:Energy} and the laboratory frame~\eq{eq:E:j} as follows:
\beqn
\nu_{ml} -1 = \frac{E_{ml} - \tE_{ml}}{E_{ml}} - 1 =  - \frac{\tE_{ml}}{E_{ml}}\,.
\label{eq:nu:ml:2}
\eeqn
However it is known that $\tE E > 0$ for the MIT boundary conditions~\eq{eq:MIT:boundary} provided the physical requirement $|\Omega| R < 1$ is valid~\eq{ref:bound}, see Ref.~\cite{Ambrus:2015lfr}. Therefore Eq.~\eq{eq:nu:ml:2} implies that $\nu_{ml} -1 < 0,$ \footnote{In Ref.~\cite{Ambrus:2015lfr}, the $q_{ml} = 0$ solutions of Eq.~\eq{eq:jj}, which may potentially give nonzero contributions to Eq.~\eq{eq:V:rot:3}, have not been discussed explicitly. We checked that there is no normalizable solution with vanishing $q_{ml}$ that are satisfying the MIT boundary conditions~\eq{eq:MIT:boundary}.} and we come to the conclusion that at zero temperature the rotational contribution to the thermodynamic potential is exactly zero. In other words {\emph{the cold vacuum does not rotate}}.

The rotational response of the system may be qualified in terms of the angular momentum and the moment of inertia that will be discussed in details in Sect.~\ref{sec:angular}. Here we notice that the insensitivity of the condensed medium to rotation in the zero temperature limit agrees well with the result obtained in the cold atom systems where it was shown that the condensed (superfluid) fraction of the system does not contribute to the moment of inertia~\cite{ref:Stringari}.

\section{Properties of a rotating fermionic matter at finite temperature}
\label{sec:rotating:finite:T}

\subsection{Phase structure and chiral symmetry breaking}
\label{sec:phase}

In this Section we discuss properties of rotating fermionic matter at finite temperature. As an example, we take large values of the coupling constant $G = 42/\Lambda^2$ and the cylinder radius $R = 20\,\Lambda^{-1}$. According to Fig.~\ref{fig:sigma:T0}(a) at these values the $T=0$ ground state is quite close to the thermodynamic limit. 

\begin{figure}[!thb]
\begin{center}
\begin{tabular}{cc}
\includegraphics[scale=0.5,clip=true]{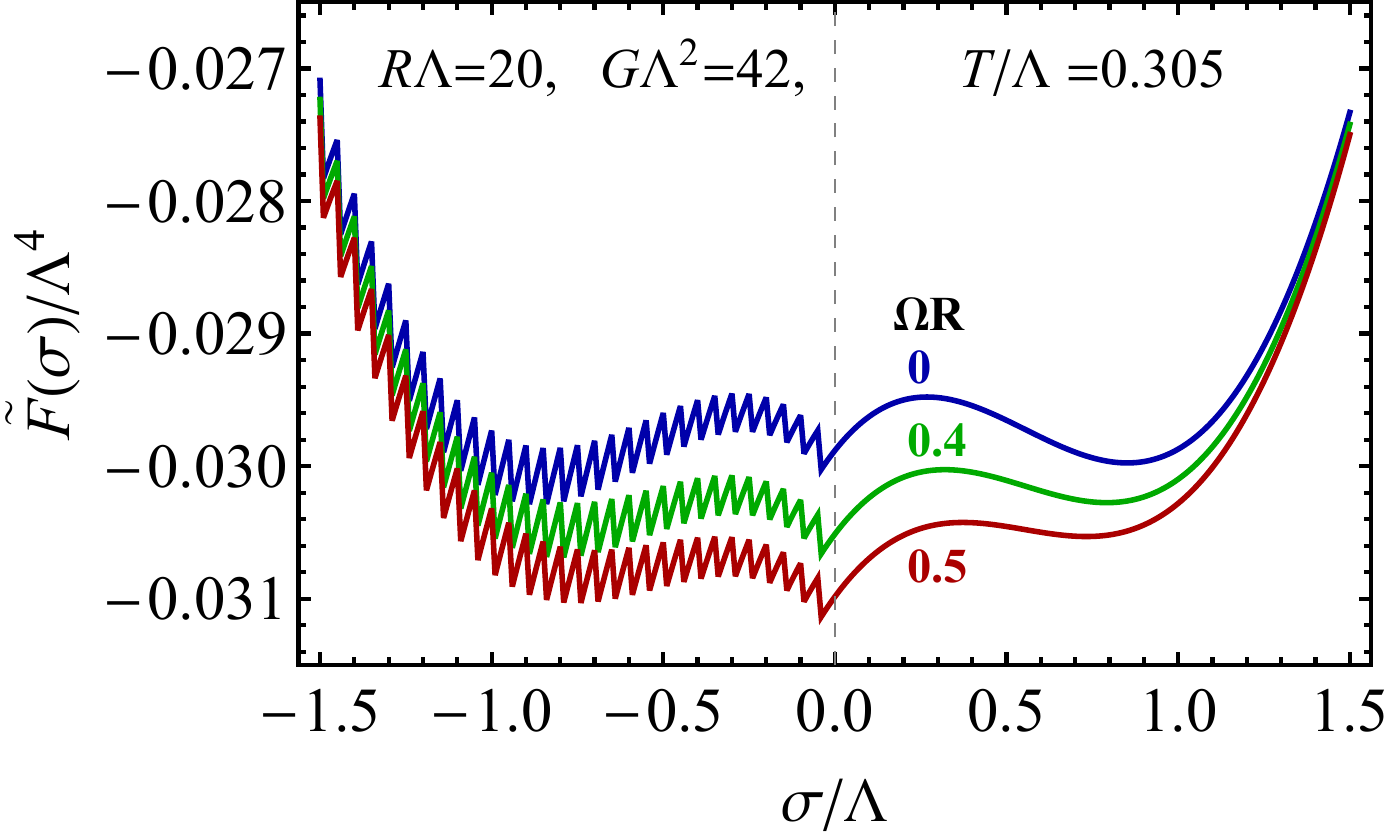} & 
\includegraphics[scale=0.5,clip=true]{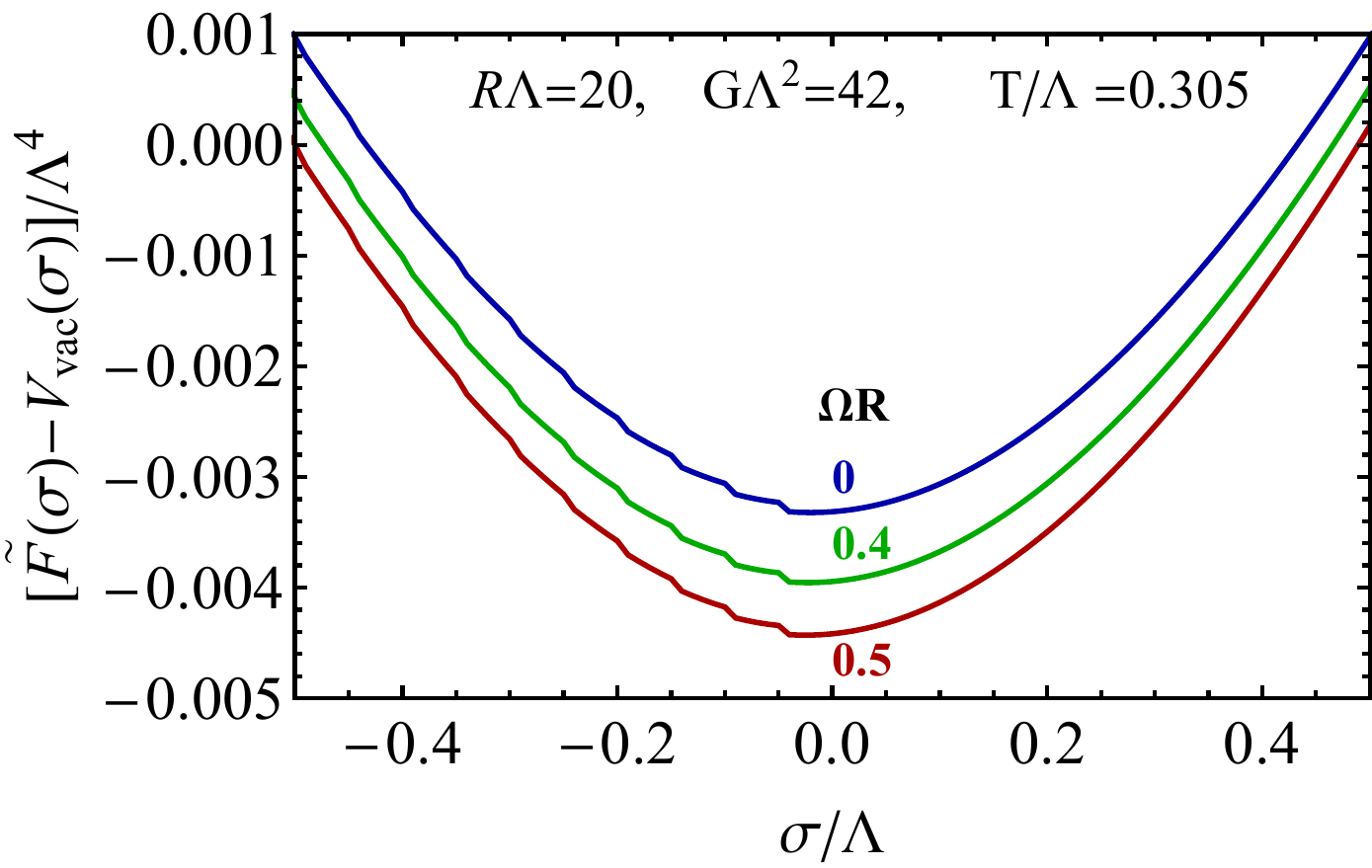} \\
(a) & (b)
\end{tabular}
\end{center}
\caption{(a) Free energy (defined in the corotating frame) for a rotating fermion matter inside the cylinder of the radius $R = 20\,\Lambda^{-1}$ at the temperature $T = 0.305 \Lambda$ and at the coupling $G = 42/\Lambda^2$ for a set of angular frequencies $\Omega$. (b) The same but for the free energy excluding the regularized vacuum part.}
\label{fig:sigma:T}
\end{figure}

In Fig.~\ref{fig:sigma:T}(a) we show the free energy~\eq{eq:F:free:energy} as the function of the field $\sigma$ for a set of the angular frequencies $\Omega$ at fixed temperature $T = 0.305 \Lambda$. As expected, the left part (negative $\sigma$) of the potential exhibits a saw-like behavior while the right part (positive~$\sigma$) is a smooth function of $\sigma$. The ground-state value of the condensate is determined by a minimum of the free energy $F(\sigma)$ which -- similarly to the zero-temperature case -- takes place at negative $\sigma$'s for all studied values of the angular frequencies $\Omega$. At this temperature the minimum of the free energy at $\Omega = 0$ corresponds to the chirally broken phase characterized by a large negative $\sigma$, the minimum of the free energy at $\Omega R = 0.5$ is very close to zero (the chirally restored phase) while at the frequency $\Omega R = 0.4$ the two minima are almost degenerate. Thus we conclude that the rotation in a finite cylinder leads to a restoration of the chiral symmetry. 

At this point we would like to stress that the saw-like behavior is not a direct consequence of the regularization of the vacuum part of the energy as the irregularities are also naturally present at the thermal part of the free energy. In order to demonstrate the latter, we show in  Fig.~\ref{fig:sigma:T}(b) the free energy~\eq{eq:F:free:energy} with the vacuum term subtracted so that only the only thermal part and the smooth $\sigma^2/2G$ term contribute to this quantity. The thermal potential leads to the saw-like dependence of the free energy although a bit less pronounced compared to the vacuum part.

\begin{figure}[!thb]
\begin{center}
\begin{tabular}{cc}
\includegraphics[scale=0.48,clip=true]{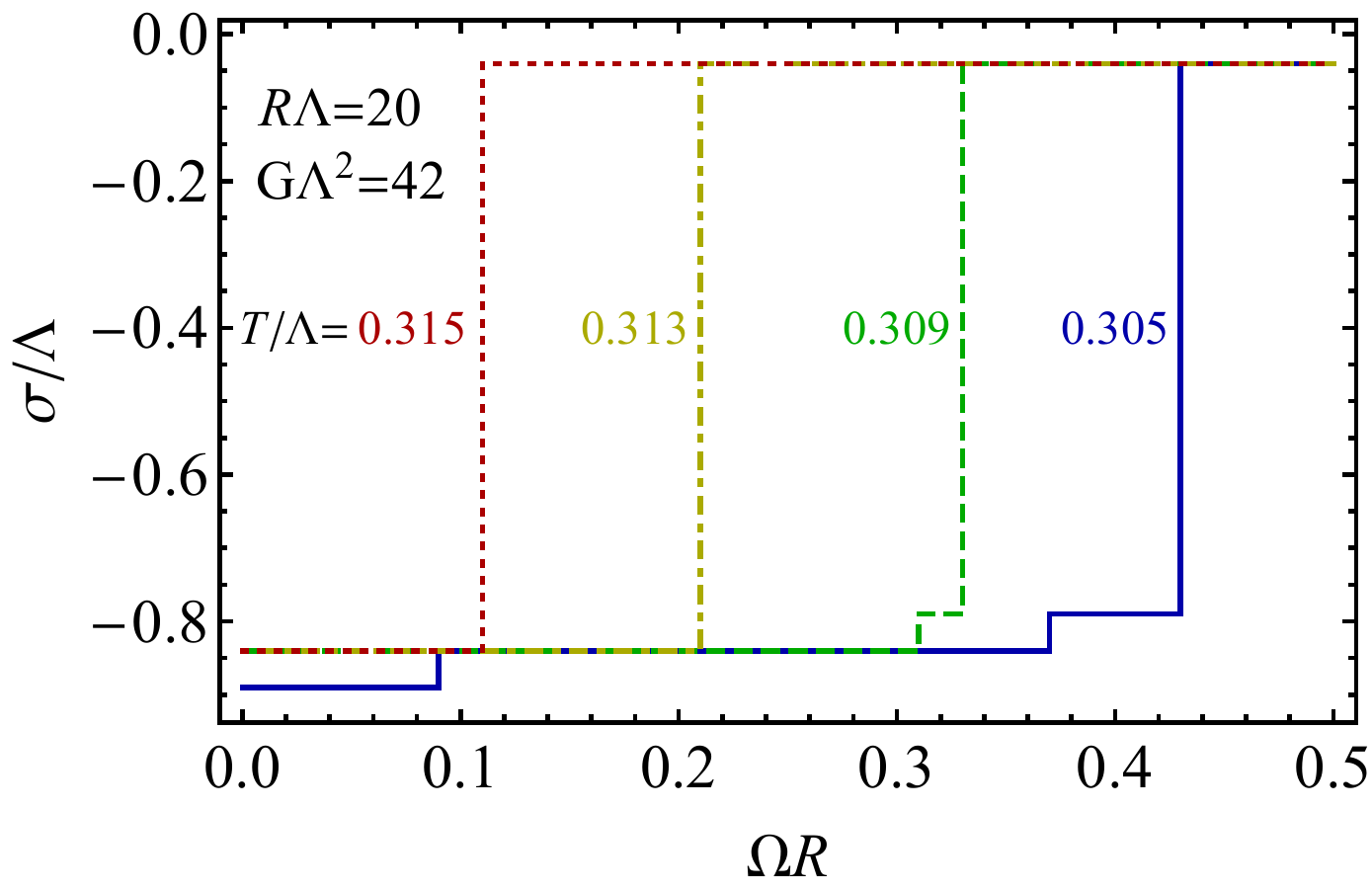} &
\includegraphics[scale=0.49,clip=true]{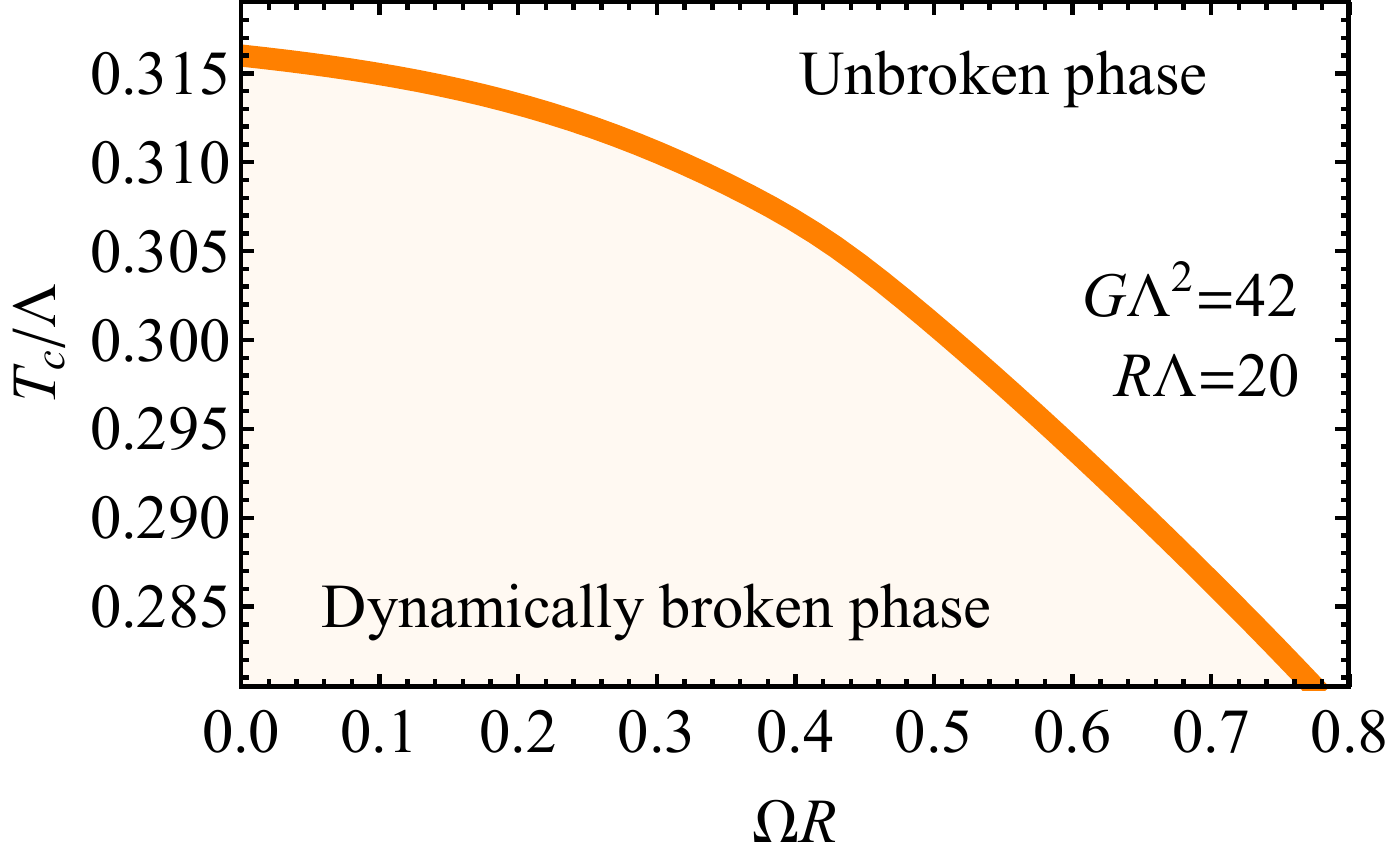} \\
(a) & (b)
\end{tabular}
\end{center}
\caption{(a) Condensate (mass gap) $\sigma$ for the fermionic matter which is rigidly rotating with the angular frequency $\Omega$ inside the cylinder of the radius $R = 20\,\Lambda^{-1}$ at the coupling $G = 42/\Lambda^2$ for a set of fixed temperatures~$T$. (b) The phase diagram of the rotating fermionic matter in the $T$-$\Omega$ plane.}
\label{fig:phase:T}
\end{figure}

In Fig.~\ref{fig:phase:T}(a) we show the ground-state condensate (mass gap) $\sigma$ as the function of the angular frequency $\Omega$ at a set of fixed temperatures $T$. The chiral symmetry breaking in the rotating environment has a few interesting features:

\begin{enumerate}

\item The rotation restores (dynamically) broken chiral symmetry at all temperatures.

\item The restoration of the chiral symmetry happens abruptly, and generally in a series of discontinuous steps. At certain temperatures the restoration happens in smaller steps, but is always discontinuous. Therefore the restoration transition is of the first order.

\item 
The rotation-induced steplike changes in the condensate are similar to the conventional Shubnikov--de Haas oscillations that are caused by external magnetic field. As we have already discussed, this similarity is a direct consequence of the finite-sized geometry of the rigidly rotating system. It is also important to stress that the rotational SdH steps occur in the absence of both external magnetic field and Fermi surface contrary to the conventional magnetic SdH oscillations.

\item The chirally broken and chirally restored values of the condensate -- respectively, before and after the step-like transition --  are almost [apart from a small step-like discontinuities in the broken region, visible in in Fig.~\ref{fig:phase:T}(a)] independent of the angular frequency $\Omega$ and temperature $T$. The critical temperature of the transition itself does depend on frequency, $T_c = T_c(\Omega)$, see Fig.~\ref{fig:phase:T}(b) below. In the restored phase the value of the condensate is small but it is still nonzero due to the explicit breaking of the chiral symmetry by the MIT boundary conditions. The residual condensate corresponds to the largest (rightmost) "tooth" in the saw-like negative part of the free energy, Fig.~\ref{fig:sigma:T}(a), the position of which is independent of the angular frequency~$\Omega$.

\end{enumerate}

We would like to remind that the negative value of the ground state condensate $\sigma$ is not a universal feature as it is a direct consequence of the MIT boundary conditions~\eq{eq:MIT:boundary} which break the chiral symmetry explicitly. For example, the so-called chiral analogue of the MIT boundary condition~\eq{eq:chiral:boundary} would lead to the exactly same results but with the sign flip in the condensate $\sigma \to - \sigma$. 

In Fig.~\ref{fig:phase:T}(b) we present the phase diagram of the rotating fermionic matter in the $T-\Omega$ plane. We find that the critical temperature of the chiral symmetry restoration $T_c$ drops down with increase of the angular frequency $\Omega$. 
The behavior of the condensate suggests that the transition in between the chirally broken and chirally restored phases is of the first order regardless of temperature because the discontinuity in the condensate (mass gap) $\sigma$ at the transition does not depend on $T$. We will show below that despite of the $T$--independent discontinuity in the mass gap $\sigma$, the free energy behavior softens across the transition as temperature increases. Notice that as shown in Fig.~\ref{fig:phase:T}(a), our finite-geometry result for the condensate qualitatively disagrees with the result of Ref.~\cite{Jiang:2016wvv} which was obtained in rotating but unbounded space, where the behavior of the condensate depends strongly on temperature $T$. However, as shown in Fig.~\ref{fig:phase:T}(b), the qualitative form of the phase diagram obtained in the bounded space agrees with the result of Ref.~\cite{Jiang:2016wvv}.

\subsection{Angular momentum and moment of inertia}
\label{sec:angular}

A mechanical response of a physical system with respect to a global rotation can be quantified in terms of the angular momentum $\bs L$ and an associated moment of inertia $I$ of the system as a whole. Since the rotation leads to the chiral symmetry restoration of the fermionic liquid in seemingly first order phase transition, one may naturally expect that both the angular momentum and the moment of inertia of the rotating fermionic medium may exhibit a non-analytic behavior across the phase transition. 

We have already seen that for each fermionic level the eigenenergy in the corotating frame ${\tE}$ is expressed via the eigenenergy in the laboratory frame $E$ and the corresponding angular momentum~$\mu_m$ with the help of a linear relation~\eq{eq:Energy}. Averaging Eq.~\eq{eq:Energy} over the whole thermodynamical ensemble in the rotating frame we get the following relation between the corresponding thermodynamic quantities~\cite{ref:LL5}:
\beqn
{\tE} = E - {\bs L} {\bs \Omega}\,,
\label{eq:E:rotating}
\eeqn
which imply that
\beqn
{\bs L} = - {\left( \frac{\partial {\tilde E}}{\partial {\bs \Omega}} \right)}_{S}\,,
\eeqn
and, consequently, $d {\tE} = T d S - {\bs L} d {\bs \Omega}$. For the free energy in the rotating frame 
\beqn
{\widetilde F} = {\tE} - T S\,,
\label{eq:F:rotating}
\eeqn
one then gets $d {\widetilde F} = - S d T - {\bs L} d {\bs \Omega}$, which immediately leads to the following useful identieis~\cite{ref:LL5}:
\beqn
{\bs L} = - {\left( \frac{\partial {\tilde F}}{\partial {\bs \Omega}} \right)}_{T}\,,
\qquad
S = - {\left( \frac{\partial {\tilde F}}{\partial T} \right)}_{{\bs \Omega}}\,.
\label{eq:L:via:tildeF}
\eeqn
For completeness we notice that the free energy in the laboratory frame, $F = E - T S$ the same considerations lead to the relation $d F = - S d T + {\bs \Omega} d {\bs L} $ which imply that independent variables in this case are the temperature $T$ and the angular momentum ${\bs L}$. One gets, consequently, the following relation associated with the angular momentum,
\beqn
{\bs \Omega} = {\left( \frac{\partial F}{\partial {\bs L}} \right)}_{T}\,,
\label{eq:Omega:via:F}
\eeqn
which is much less useful for our purposes compared to Eq.~\eq{eq:L:via:tildeF} that involves the free energy in the corotating frame. 

The moment of inertia $I = I(\Omega)$ may naturally be defined as a linear response of the angular momentum $\bs L$ of the system with respect to the angular velocity $\bs \Omega$, Ref.~\cite{ref:Stringari}:
\beqn
{\bs L} = I(\Omega) {\bs \Omega}\,.
\eeqn
Then, according to Eq.~\eq{eq:L:via:tildeF} the moment of inertia can be expressed as follows:
\beqn
I(\Omega) \equiv \frac{L(\Omega)}{\Omega} = - \frac{1}{\Omega} {\left( \frac{\partial {\tilde F}}{\partial \Omega} \right)}_{T}\,,
\label{eq:moment:inertia}
\eeqn
where we reduced the vector notations ${\bs \Omega} = \Omega {\bs e} $ by setting that the rotation goes about a fixed axis ${\bs e}$.

\begin{figure}[!thb]
\begin{center}
\includegraphics[scale=0.48,clip=true]{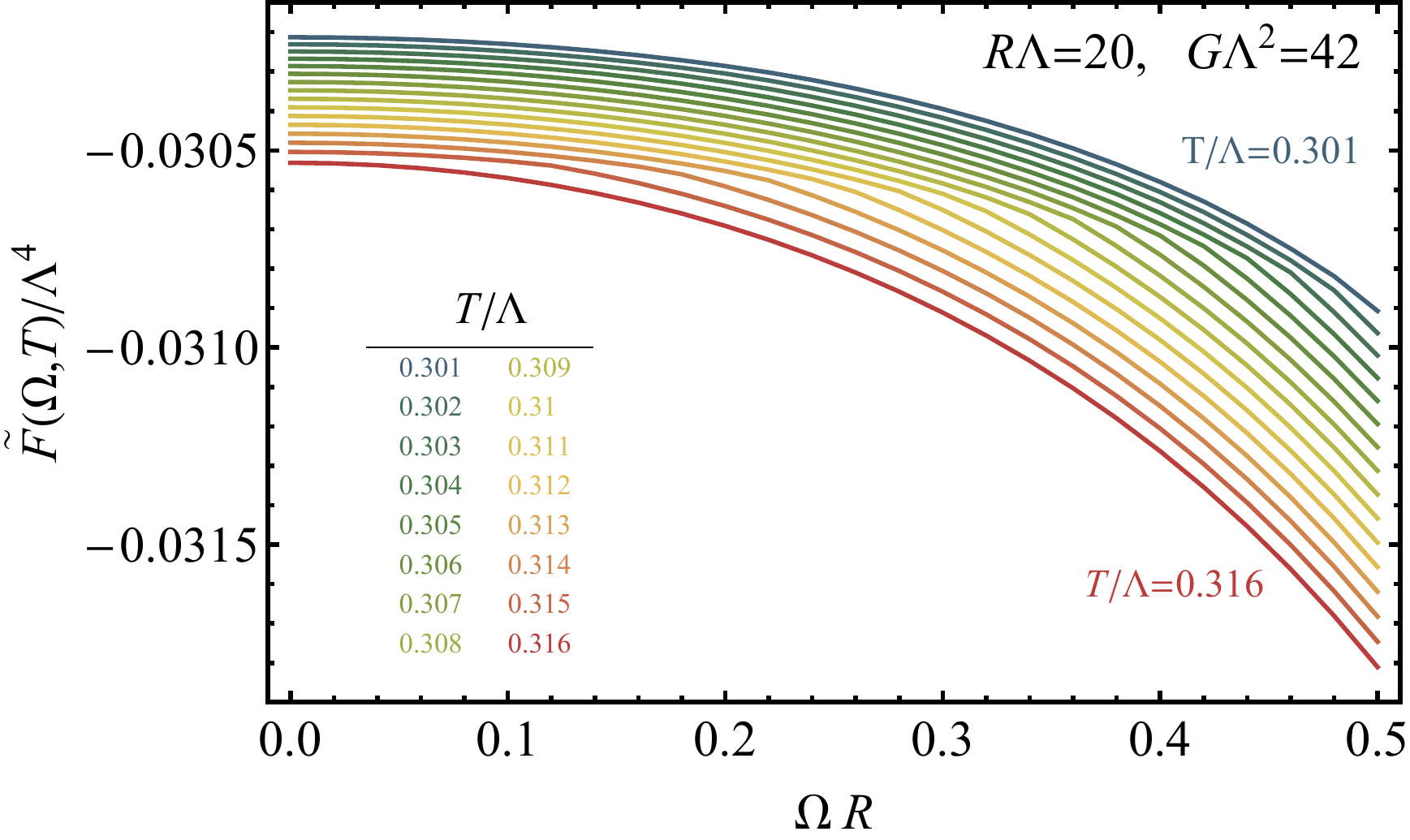}
\end{center}
\caption{The ground-state free energy~\eq{eq:F:free:energy} determined in the rotating frame vs the angular frequency $\Omega$ for the fixed coupling $G = 42 \Lambda^{-2}$, fixed radius $R = 20 \Lambda^{-1}$ and various temperatures $T$ ranging from $T=0.301 \Lambda$ (the upper curve) till $T=0.316 \Lambda$ (the lower curve). A knee-lie structure at the phase transition from the broken phase (lower $\Omega$) to the restored phase (higher $\Omega$) is clearly seen for each temperature.}
\label{fig:free:energy:Omega}
\end{figure}

In Fig.~\ref{fig:free:energy:Omega} we show the ground-state free energy in the rotating frame as a function of the angular frequency $\Omega$ at a fixed set of temperatures. The temperature range is chosen to cover the region of the phase transition shown previously in Fig.~\ref{fig:phase:T}(b). As above, we fix the coupling constant $G = 42 \Lambda^{-2}$ and radius $R = 20 \Lambda^{-1}$ of the fermionic ensemble.

Notice that the free energy shown in Fig.~\ref{fig:free:energy:Omega} is evaluated at the ground state $\sigma$ which is determined as a minimum of the free energy~\eq{eq:F:free:energy} for a given fixed set of parameters $\Omega$ and $T$ (therefore we refer to this quantity as the ``ground-state free energy''). The phase transition is clearly visible as a change of the slope which appears at distinct angular frequencies $\Omega$ for different  temperatures $T$. The changes in the slope indicate that the rotational properties of the fermionic medium are different in chirally broken and chirally restored phases. 

\begin{figure}[!thb]
\begin{center}
\begin{tabular}{cc}
\includegraphics[scale=0.415,clip=true]{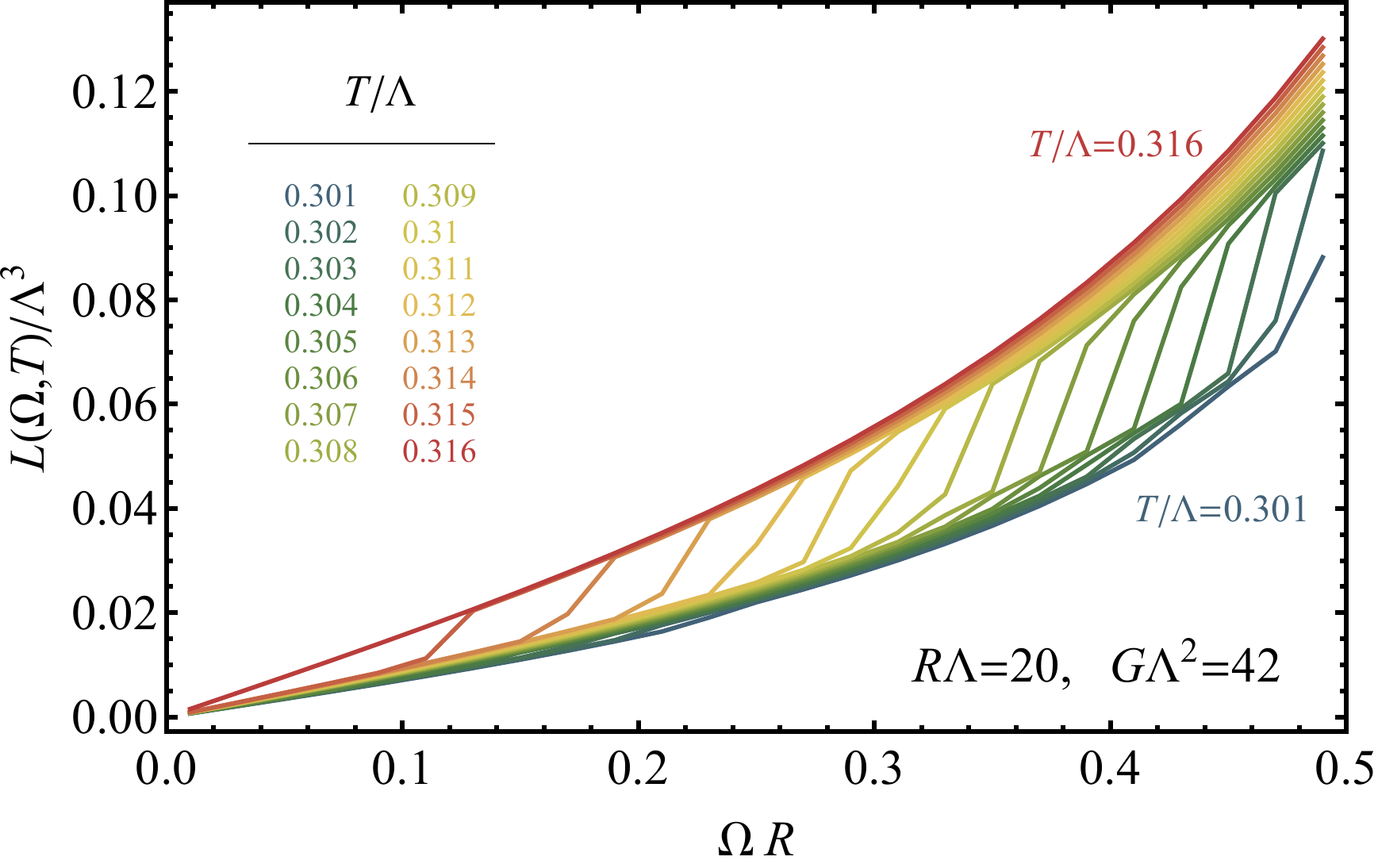} &
\includegraphics[scale=0.4,clip=true]{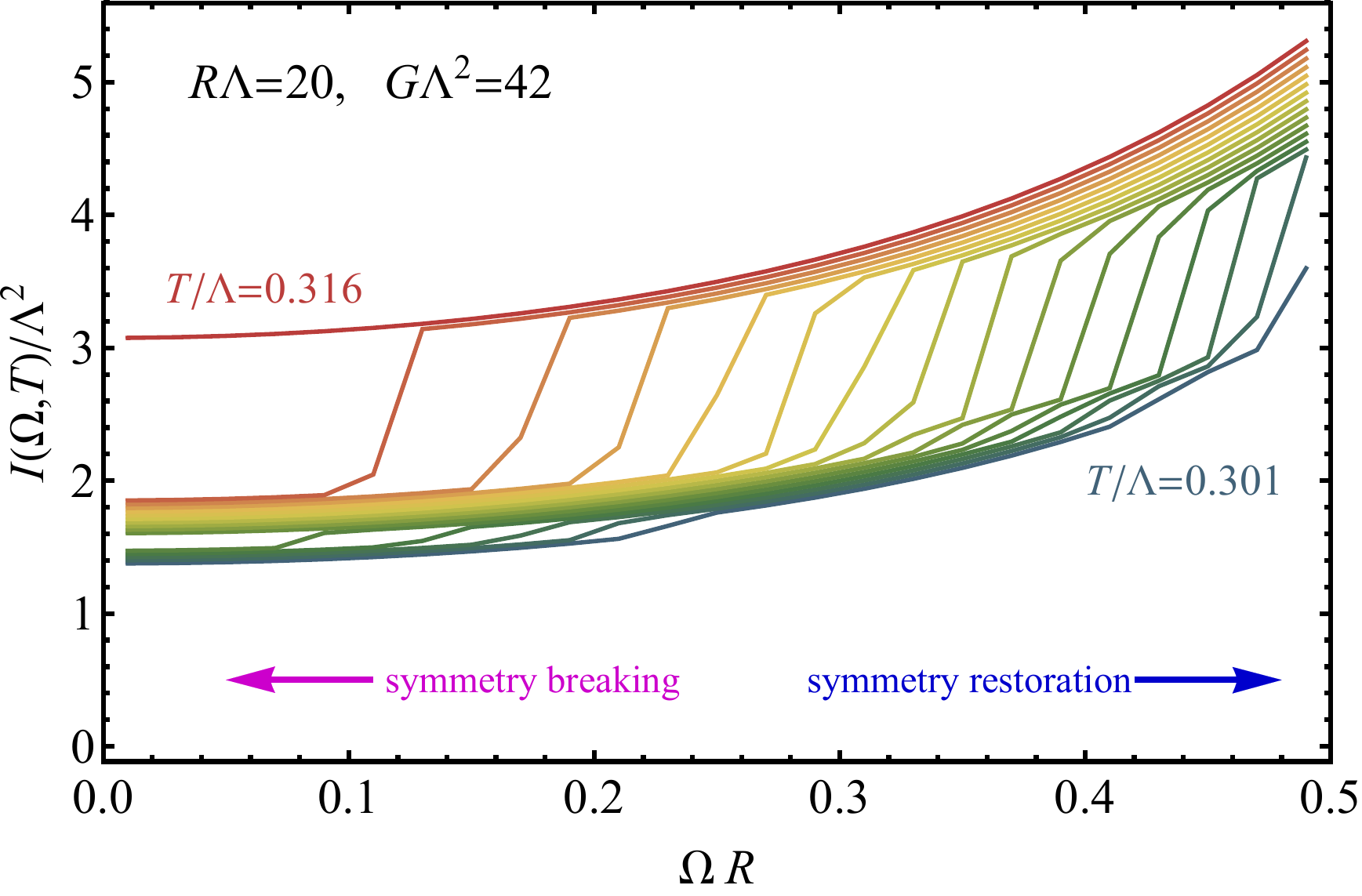} \\
(a) & (b)
\end{tabular}
\end{center}
\caption{(a) The angular momentum~\eq{eq:L:via:tildeF} and (b) the moment of inertia~\eq{eq:moment:inertia} of the rotating fermionic matter across the phase transition line. The parameters are the same as in Fig.~\ref{fig:free:energy:Omega}.}
\label{fig:angular:mometum:Omega}
\end{figure}

The angular momentum of the rotating fermions~\eq{eq:L:via:tildeF} is shown in Figs.~\ref{fig:angular:mometum:Omega}(a). The angular momentum in the chirally broken phase is lower than the angular momentum in the chirally restored phase. The two phases are separated by a sharp increase of the angular momentum indicating, counterintuitively, that the chirally restored, gapless phase is more rotationally ``massive'' as compared to the chirally broken, gapped phase. 

The same observation is confirmed by the behavior of the moment of inertia~\eq{eq:moment:inertia} which is shown in Figs.~\ref{fig:angular:mometum:Omega}(b) for the same set of parameters. The moment of inertia $I(\Omega)$ of the fermionic matter in the chirally broken phase (lower $\Omega$) is lower compared to the rotating matter in the chirally restored phase (higher $\Omega$). The moment of inertia is not a constant function of the frequency as it growth up in both phases, and exhibits a rapid increase in the transition region.

\subsection{Energy and entropy of rotating fermions}
\label{sec:energy}

\begin{figure}[!thb]
\begin{center}
\begin{tabular}{cc}
\includegraphics[scale=0.415,clip=true]{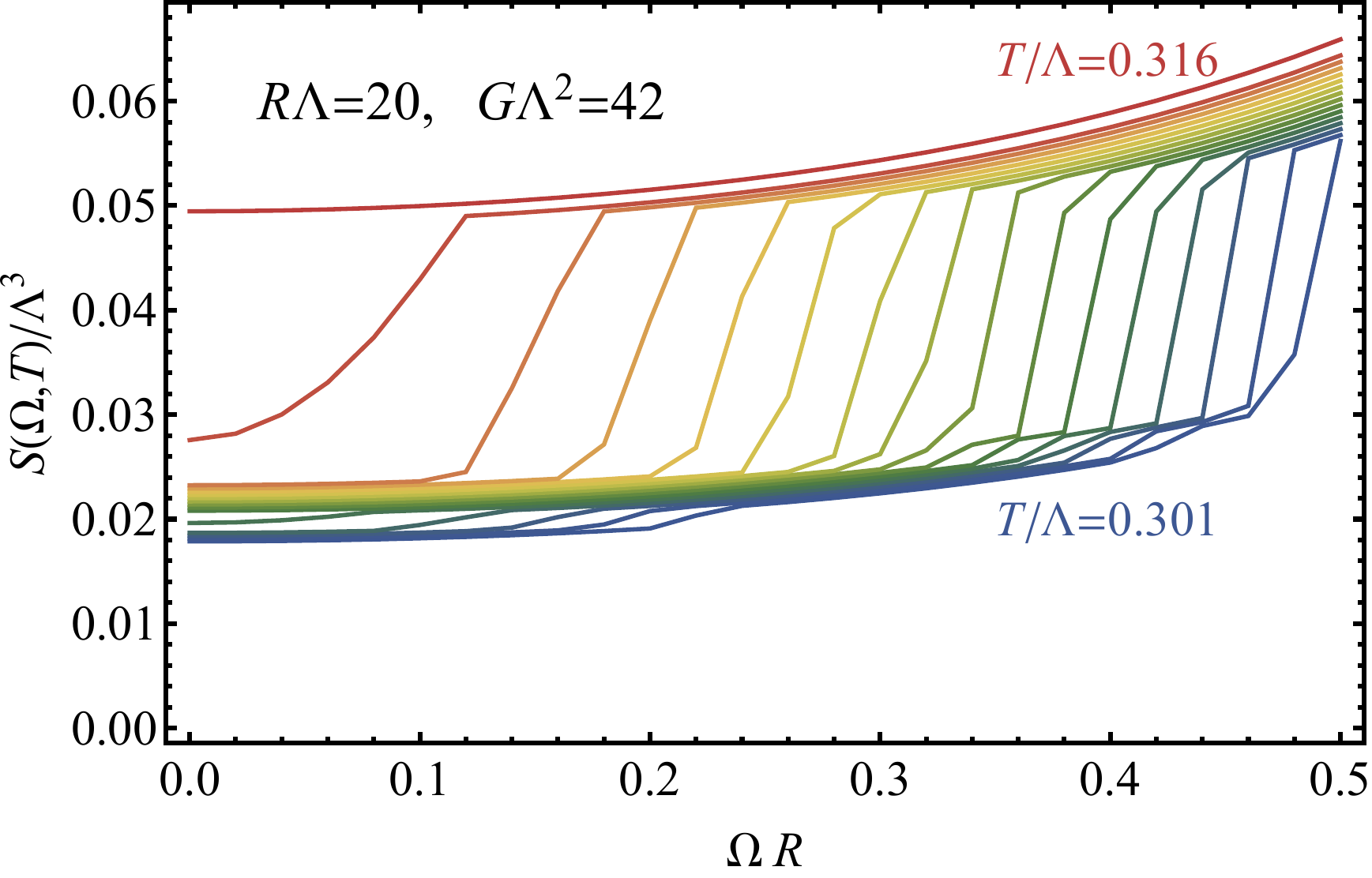} &
\includegraphics[scale=0.415,clip=true]{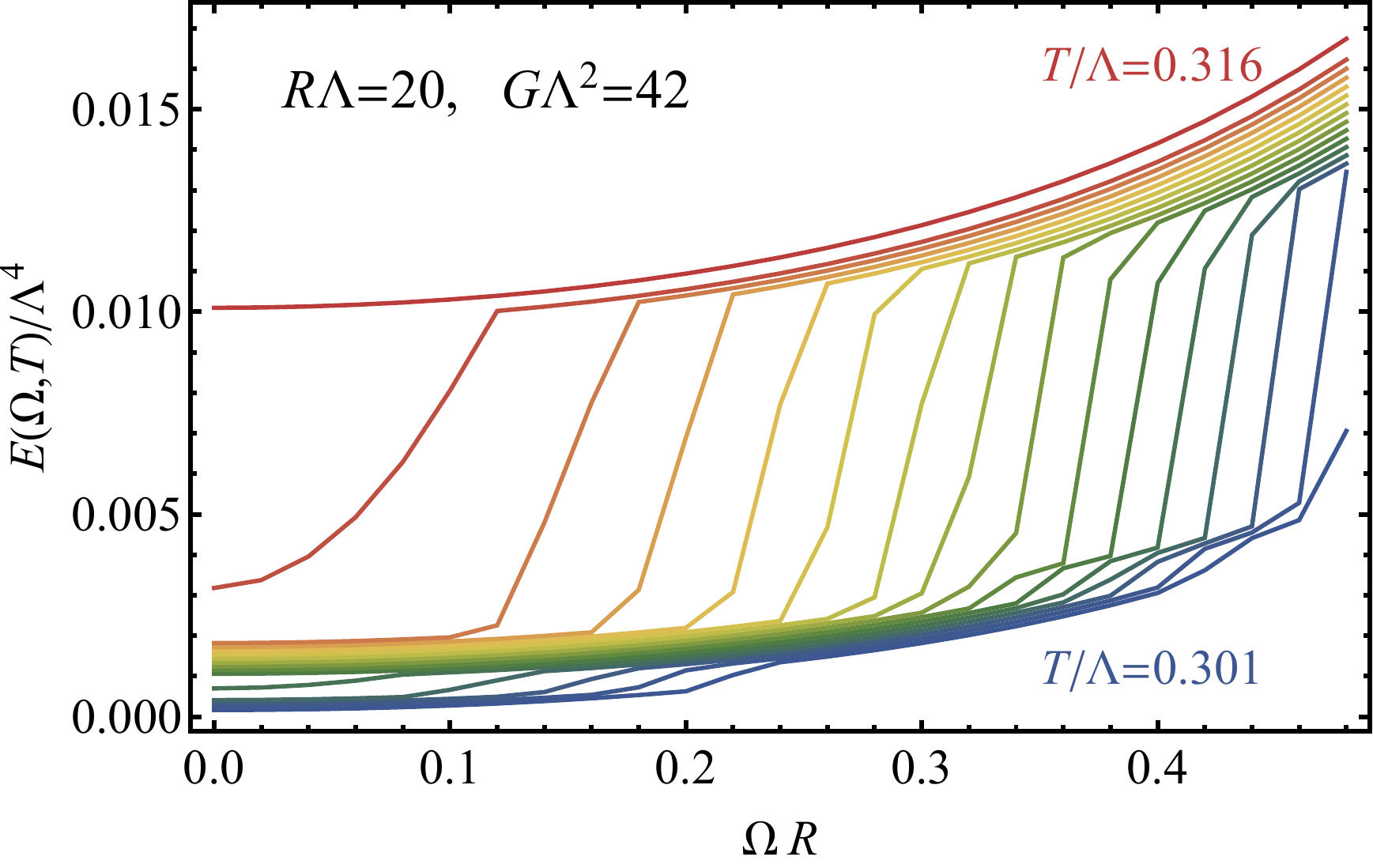} \\
(a) & (b)
\end{tabular}
\end{center}
\caption{(a) Entropy density~\eq{eq:L:via:tildeF} and (b) energy density in the laboratory frame~\eq{eq:energy:laboratory} as the function of rotation frequency~$\Omega$ of the fermionic matter. The parameters are the same as in Fig.~\ref{fig:free:energy:Omega}.}
\label{fig:entropy:energy:Omega}
\end{figure}

According to Eqs.~\eq{eq:E:rotating} and \eq{eq:F:rotating} the energy density in the laboratory frame is given by the following formula:
\beqn
E = {\widetilde F} + {\bs L} {\bs \Omega} + T S\,,
\label{eq:energy:laboratory}
\eeqn
where the angular momentum ${\bs L}$ and the entropy $S$ can be obtained with the help of Eq.~\eq{eq:L:via:tildeF}. We plot the entropy and energy densities in the laboratory frame in Fig.~\ref{fig:entropy:energy:Omega}.

Both entropy and energy (as determined in the laboratory frame) of the rotating fermions experience a visible change with increase of the rotational frequency $\Omega$ as the system passes from the chirally broken region to the chirally restored region. The entropy and energy are smaller at the chirally broken region at low $\Omega$ compared to their values in chirally restored region the at higher $\Omega$. Moreover, the transition between these regions becomes substantially smoother with increase of temperature. The rotational SdH steplike features are also seen in Fig.~\ref{fig:entropy:energy:Omega} at low temperatures.

\section{Discussion and Conclusions}

In our paper we concentrated on properties of rotating systems of interacting fermions in the framework of the Nambu--Jona-Lasinio model. Before summarizing the results of our studies we would like to stress that the problem of relativistic rigid rotation should always be studied in a spatially bounded physical volume and thus should depend on specifics of conditions imposed on fermion fields at the boundaries of the volume.

Indeed, the velocity of any particle in a rotating system should not exceed the speed of light. Therefore, a rigid rotation of relativistic matter in thermodynamic equilibrium must always be considered in a volume which is bounded in directions perpendicular to the axis of rotation. The finite geometry implies that physical properties of the system should generally depend on conditions that are imposed on the fields at the boundaries of the volume. Thus, we come to the conclusion that physical systems in rotation should always be considered in conjunction with appropriate boundary conditions. Moreover, a faster rotation implies a smaller size of the system in the direction transverse to the rotation axis. The latter inevitably leads to a stronger dependence of properties of the system on the boundary conditions. One may look to this problem from a different perspective: the boundary conditions are becoming increasingly irrelevant for larger volumes for which, however, the rotation must be constrained to smaller angular frequencies. Thus, the properties of any rigidly rotating system should always be formulated in finite geometries with appropriate boundary conditions.

In our studies we considered the system of rotating fermions bounded by cylindrical shell with the MIT boundary conditions that physically confine the fermions inside the cylinder of the fixed radius $R$.  Our main results can be summarized as follows:
\begin{enumerate}

\item {\bf Boundaries are very important.} The MIT boundary conditions affect strongly the phase structure of interacting fermions in cylinder. The finite boundaries brake the chiral symmetry explicitly in the interior of the cylinders with the small radius $R \lesssim (2\dots 3) \Lambda^{-1}$. They lead to the restoration of the chiral symmetry in the moderately-sized cylinders thus not allowing the dynamical chiral symmetry to occur at radii $ (2\dots 3) \Lambda^{-1} \lesssim R < R_c(G)$. The dynamical condensation allowed at relatively large radius of the cylinder, $R > R_c(G)$, where the critical condensation radius $R_c(R)$ is a rising function of the coupling $G$. A typical behavior of chiral condensate as the function of the radius of the cylinder is shown in Fig.~\ref{fig:sigma:T0}(a). The zero-temperature phase diagram in the radius--coupling constant plane is shown in Fig.~\ref{fig:phase}

\item {\bf Boundary conditions break explicitly the reflection symmetry of the mass gap}, $\sigma \to -\sigma$, in general. For the MIT boundary conditions ~\eq{eq:MIT:boundary} the mean-field value of the condensate is negative, $\sigma < 0$, while for the chiral MIT boundary conditions~\eq{eq:chiral:boundary} the condensate takes positive values, $\sigma > 0$. The two boundary conditions are related to each other by the chiral transformation~\eq{eq:chiral:transformations} with the chiral angle  $\theta = \pi/2$. In the infinite volume, $R \to \infty$, the reflection symmetry $\sigma \to -\sigma$ is restored, Fig.~\ref{fig:Vvac}.

\item {\bf Rotational Shubnikov--de Haas-like effect}. The presence of the boundary leads to specific steplike irregularities of the chiral condensate and other quantities as functions of coupling constant $G$, Fig.~\ref{fig:F:sigma:vacuum}(b), temperature $T$ and radius $R$, Fig.~\ref{fig:sigma:T0}(a), and, most importantly, in angular frequency $\Omega$, Fig.~\ref{fig:phase:T}(a). We argued that these steplike features occur due to discreteness of the energy levels of fermions which result, for example, in well-pronounced steplike behavior of free energy both at zero temperature, Fig.~\ref{fig:F:sigma:vacuum} and at finite temperature, Fig.~\ref{fig:sigma:T}. The predicted steplike features have the same nature as the Shubnikov--de Haas oscillations in the conductivity of a material with the crucial, however, difference that they occur in the absence of both external magnetic field and Fermi surface.

\item {\bf Cold vacuum cannot rotate}. The vacuum at zero temperature vacuum is insensitive to rotation, Sect.~\ref{sec:cold:vacuum}, in agreement with a similar result obtained in nonrelativistic bosonic cold atom systems~\cite{ref:Stringari}. The same property has been also noted recently for relativistic fermionic systems in Ref.~\cite{Ebihara:2016fwa}. 

\item {\bf Rotation cannot be associated with fictitious magnetic field.} We have provided the arguments -- which are based on the energy level structure, associated density of states and levels' degeneracy -- that the rigid rotation of a relativistic system cannot be associated with an external magnetic field thus supporting a similar statement made recently in Ref.~\cite{Chen:2015hfc}.

\item {\bf At finite temperature the rotation leads to restoration of spontaneously broken chiral symmetry.} The phase diagram in the angular frequency-temperature plane for a cylinder with a finite radius is shown in Fig.~\ref{fig:phase:T}(b). It agrees qualitatively with calculation of Ref.~\cite{Jiang:2016wvv} made in an unbounded space.

\item {\bf Softening of the transition strength with increase temperature.} As the temperature increases, the critical angular frequency decreases and the transition becomes softer according to behavior of the free energy density in rotating frame, Fig.~\ref{fig:free:energy:Omega} (the softening is well seen in entropy density and energy density in the laboratory frame, Fig.~\ref{fig:entropy:energy:Omega}). However in our approach -- contrary to the calculations of Ref.~\cite{Jiang:2016wvv}, where the boundary effects are not taken into account -- the discontinuity of the mass gap across the phase transition does not depend on the angular frequency~$\Omega$ if one takes boundary conditions into account, Fig.~\ref{fig:phase:T}(a).

\item {\bf Inequivalence of angular momentum and moment of inertia in different phases.} At fixed temperature the fermion matter in the chirally restored (higher $\Omega$) region has a higher angular momentum and higher moment of inertia compared to the ones in the chirally broken (lower $\Omega$) region, Fig.~\ref{fig:entropy:energy:Omega}. In the transition region both quantities grow significantly as the angular frequency~$\Omega$ is increasing.

\end{enumerate}

\acknowledgments 

The work of S.~G. was supported by a grant from La Region Centre (France).


\begin{thebibliography}{99}
\bibitem{ref:HIC:1}
  L.~P.~Csernai, V.~K.~Magas and D.~J.~Wang,
``Flow Vorticity in Peripheral High Energy Heavy Ion Collisions,''
  Phys.\ Rev.\ C {\bf 87}, no. 3, 034906 (2013)
  [arXiv:1302.5310 [nucl-th]].

\bibitem{ref:HIC:2}
  F.~Becattini {\it et al.},
  ``A study of vorticity formation in high energy nuclear collisions,''
  Eur.\ Phys.\ J.\ C {\bf 75}, no. 9, 406 (2015)
  [arXiv:1501.04468 [nucl-th]].

\bibitem{ref:HIC:3}
    Y.~Jiang, Z.~W.~Lin and J.~Liao,
  ``Rotating quark-gluon plasma in relativistic heavy ion collisions,''
  Phys.\ Rev.\ C {\bf 94}, no. 4, 044910 (2016)
  [arXiv:1602.06580 [hep-ph]].

\bibitem{ref:HIC:4}
    W.~T.~Deng and X.~G.~Huang,
  ``Vorticity in Heavy-Ion Collisions,''
  Phys.\ Rev.\ C {\bf 93}, no. 6, 064907 (2016)
  [arXiv:1603.06117 [nucl-th]].

\bibitem{ref:CVE:1} 
  D.~T.~Son and A.~R.~Zhitnitsky,
``Quantum anomalies in dense matter,''
  Phys.\ Rev.\ D {\bf 70}, 074018 (2004)
  [hep-ph/0405216].

 \bibitem{ref:CVE:2} 
 D.~T.~Son and P.~Surowka,
  ``Hydrodynamics with Triangle Anomalies,''
  Phys.\ Rev.\ Lett.\  {\bf 103}, 191601 (2009)
  [arXiv:0906.5044 [hep-th]].

\bibitem{ref:Vilenkin}
  A.~Vilenkin,
  ``Parity Violating Currents in Thermal Radiation,''
  Phys.\ Lett.\  {\bf 80B}, 150 (1978);
 ``Macroscopic Parity Violating Effects: Neutrino Fluxes From Rotating Black Holes And In Rotating Thermal Radiation,''
  Phys.\ Rev.\ D {\bf 20}, 1807 (1979);
``Quantum Field Theory At Finite Temperature In A Rotating System,''
  Phys.\ Rev.\ D {\bf 21}, 2260 (1980).
 
\bibitem{ref:Weyl:1} 
  G.~Basar, D.~E.~Kharzeev and H.~U.~Yee,
  ``Triangle anomaly in Weyl semimetals,''
  Phys.\ Rev.\ B {\bf 89}, no. 3, 035142 (2014)
  [arXiv:1305.6338 [hep-th]].

\bibitem{ref:Weyl:2} 
  K.~Landsteiner,
``Anomalous transport of Weyl fermions in Weyl semimetals,''
  Phys.\ Rev.\ B {\bf 89}, no. 7, 075124 (2014)
  [arXiv:1306.4932 [hep-th]].

\bibitem{ref:Weyl:3} 
  M.~N.~Chernodub, A.~Cortijo, A.~G.~Grushin, K.~Landsteiner and M.~A.~H.~Vozmediano,
  ``Condensed matter realization of the axial magnetic effect,''
  Phys.\ Rev.\ B {\bf 89}, no. 8, 081407 (2014)
  [arXiv:1311.0878 [hep-th]].

\bibitem{Ambrus:2014uqa} 
  V.~E.~Ambru\c{s} and E.~Winstanley,
  ``Rotating quantum states,''
  Phys.\ Lett.\ B {\bf 734}, 296 (2014)
  [arXiv:1401.6388 [hep-th]].

\bibitem{Ambrus:2015lfr} 
  V.~E.~Ambru\c{s} and E.~Winstanley,
  ``Rotating fermions inside a cylindrical boundary,''
  Phys.\ Rev.\ D {\bf 93}, no. 10, 104014 (2016)
  [arXiv:1512.05239 [hep-th]].

\bibitem{Chen:2015hfc} 
  H.~L.~Chen, K.~Fukushima, X.~G.~Huang and K.~Mameda,
``Analogy between rotation and density for Dirac fermions in a magnetic field,''
  Phys.\ Rev.\ D {\bf 93}, no. 10, 104052 (2016)
  [arXiv:1512.08974 [hep-ph]].

\bibitem{Jiang:2016wvv}
  Y.~Jiang and J.~Liao,
  ``Pairing Phase Transitions of Matter under Rotation,''
  arXiv:1606.03808 [hep-ph].
  
\bibitem{Ebihara:2016fwa} 
  S.~Ebihara, K.~Fukushima and K.~Mameda,
  ``Boundary effects and gapped dispersion in rotating fermionic matter,''
  arXiv:1608.00336 [hep-ph].

\bibitem{McInnes:2014haa} 
  B.~McInnes,
  ``Angular Momentum in QGP Holography,''
  Nucl.\ Phys.\ B {\bf 887}, 246 (2014)
  [arXiv:1403.3258 [hep-th]].

\bibitem{McInnes:2015kec} 
  B.~McInnes,
  ``Inverse Magnetic/Shear Catalysis,''
  Nucl.\ Phys.\ B {\bf 906}, 40 (2016)
  [arXiv:1511.05293 [hep-th]].

\bibitem{McInnes:2016dwk} 
  B.~McInnes,
  ``A rotation/magnetism analogy for the quark--gluon plasma,''
  Nucl.\ Phys.\ B {\bf 911}, 173 (2016)
  [arXiv:1604.03669 [hep-th]].

\bibitem{Yamamoto:2013zwa} 
  A.~Yamamoto and Y.~Hirono,
  ``Lattice QCD in rotating frames,''
  Phys.\ Rev.\ Lett.\  {\bf 111}, 081601 (2013)
  [arXiv:1303.6292 [hep-lat]].

\bibitem{ref:NJL}
  Y.~Nambu and G.~Jona-Lasinio,
``Dynamical Model Of Elementary Particles Based On An Analogy With Superconductivity,''
  Phys.\ Rev.\  {\bf 124}, 246 (1961);
  Phys.\ Rev.\  {\bf 122}, 345 (1961).

\bibitem{ref:Levin} 
O. Levin, Y. Peleg and A. Peres, 
``Unruh effect for circular motion in a cavity'',
J. Phys. A {\bf 26}, 3001 (1993).

\bibitem{Davies:1996ks} 
  P.~C.~W.~Davies, T.~Dray and C.~A.~Manogue,
  ``The Rotating quantum vacuum,''
  Phys.\ Rev.\ D {\bf 53}, 4382 (1996)
  [gr-qc/9601034].

\bibitem{Miransky:2015ava} 
  V.~A.~Miransky and I.~A.~Shovkovy,
  ``Quantum field theory in a magnetic field: From quantum chromodynamics to graphene and Dirac semimetals,''
  Phys.\ Rept.\  {\bf 576}, 1 (2015)
  [arXiv:1503.00732 [hep-ph]].

\bibitem{Gorbar:2011ya} 
  E.~V.~Gorbar, V.~A.~Miransky and I.~A.~Shovkovy,
``Normal ground state of dense relativistic matter in a magnetic field,''
  Phys.\ Rev.\ D {\bf 83}, 085003 (2011)
  [arXiv:1101.4954 [hep-ph]].
  
\bibitem{ref:SdH}
B. K. Ridley, Quantum Processes in Semiconductors ( 4th Edition, Oxford University Press, 2000).
  
\bibitem{ref:LL5}
L.~D.~Landau and E.~M.~Lifshitz, ``Statistical Physics, Part 1: Volume 5'',
(But\-ter\-worth\--Hei\-ne\-mann, Oxword, 1980).

\bibitem{ref:Stringari}
S. Stringari,
``Moment of Inertia and Superfluidity of a Trapped Bose Gas'',
  Phys.\ Rev.\ Lett. {\bf 76}, 085003 (1996).


\end{thebibliography}
\end{document}